\documentclass[tightenlines,twocolumn,superscriptaddress,preprintnumbers,amssymb,nofootinbib]{revtex4-1}
\usepackage{amsmath, amsthm, amssymb, mathrsfs, bm, mathtools} 
\usepackage{graphics}
\usepackage{epsfig}
\usepackage{graphicx}
\usepackage{wrapfig}
\usepackage{color}
\usepackage{tikz}

\usepackage[hidelinks,colorlinks=true,linkcolor=blue,citecolor=blue]{hyperref}

\RequirePackage{subfigure}

\newcommand{\beq}{\begin{equation}}
\newcommand{\eeq}{\end{equation}}
\newcommand{\bqa}{\begin{eqnarray}}
\newcommand{\eqa}{\end{eqnarray}}

\newcommand{\ms}{\overline{\text{\tiny MS}}}

\newcommand{\x}{x}
\newcommand{\y}{y}

\begin{document}

\title{ Comparison of chiral limit studies in curvature mass versus on-shell renormalized  quark-meson model using ChPT}
\author{Vivek Kumar Tiwari}
\email{vivekkrt@gmail.com}
\affiliation{Department of Physics, University of Allahabad, Prayagraj, India-211002}
\date{\today}

\begin{abstract}
Consistent chiral limit has been investigated in the curvature mass parametrized quark-meson (QM) model with the quark one-loop vacuum term (QMVT) employing the infrared regularized Chiral perturbation theory (ChPT) predicted scaling of the pion,~kaon decay  constants $f_{\pi}, f_{K} $ and $M_{\eta}^2 = m_{\eta}^2 + m_{\eta^{\prime}}^2 $ when the $\pi  \ \text{and} \ K $ meson masses are reduced as one moves away from the physical point in the Columbia plot.~Comparing the QMVT model Columbia plots with the corresponding Columbia plots computed,~in the  very recent work of Ref.~\cite{vkt25} using the on-shell renormalized QM (RQM) model and the earlier work of Ref.~\cite{Resch} using functional renormalization group techniques in the extended mean field approximation of QM  (e-MFA:QM-FRG) model,~it has been estimated how  the first, second and crossover chiral transition regions in the $m_{\pi}-m_{K}$($m_{ud}-m_{s}$) and the $\mu-m_{K}$($\mu-m_{s}$) planes,~get modified by different methods of implementing the quark one-loop vacuum fluctuations in the QM model.~Since both the e-MFA:QM-FRG and the QMVT model,~use curvature meson  masses to fix the parameters and the dimensional regularization of vacuum divergences are incorporated equivalently,~the differences in their results can be attributed to different methods of approaching the chiral limit.~The first order regions in the  QMVT model while being much smaller than those in the RQM model,~have similar features but moderately smaller area than those in the e-MFA:QM-FRG study.~In going to the chiral limit,~the vacuum mass $m_{\sigma}=530$ MeV that is taken at the physical point,~does not change in the QMVT model whereas it decreases significantly in the e-MFA:QM-FRG study.~Being different from the pole mass  $m_{\sigma}=530$ MeV,~the RQM model vacuum curvature mass $m_{\sigma,c}$ increases towards the chiral limit from its minimum value at the physical point.

\end{abstract}

\maketitle
\section{Introduction}
\label{secI}

The Quantum chromodynamics (QCD) Lagrangian has  the global $SU_{\tiny{V}}(3) \times SU_{A}(3) $ symmetry when the $u,d \text{ and } s$ quarks are massless.~One gets the chiral phase transition with the light and strange condensates as  order parameter and eight massless Goldstone modes as pseudoscalar mesons when the $SU_{A}(3) $ chiral  symmetry is spontaneously broken in the low energy QCD vacuum.~Since, very small light quark mass $m_{ud}$ and small strange quark mass $m_{s}$ cause a small explicit breaking of the chiral  symmetry,~one gets three light pions and moderately heavier four kaons and one $\eta$ meson in nature.~The $U_A(1)$ axial symmetry gets explicitly broken due to the $U_A(1)$ anomaly caused instanton effects and the $\eta'$ meson becomes about 1 GeV heavy~\cite{tHooft:76prl}.~The universality \cite{rob} arguments predict that the chiral phase transition,~is first order in the massless quark chiral limit $m_{ud}=0=m_{s}$ and second order belonging to the $O(4)$ universality class if $m_{u,d}=0$, $m_{s}=\infty$ and the $U_A(1)$ anomaly is strong at the critical temperature of chiral symmetry restoring phase transition  $T_{c}$ ($\eta^{\prime}$ mass $m_{\eta^{\prime}}(T_{c})>>T_{c}$).~One gets a triple point for some finite $m_{s}=m_{s}^{TCP}$($m_{K}=m_{K}^{TCP}$) when the second order transition turns first order  in the light chiral limit $m_{u,d}=0$ ($m_{\pi}=0$).~The mass sensitivity of the order of phase transition,~the critical line and critical surface structures for zero and real/imaginary 
chemical potentials $\mu$,~are summarized in the Columbia plot \cite{columb} for which, both the continuum and lattice QCD (LQCD) studies \cite{from, saito, reinosa, fischrA, mael},~have generated a good understanding in the heavy quark mass limit.~The LQCD studies have settled that the $\mu=0$ chiral symmetry restoration  at the physical point,~is a crossover transition that occurs at the pseudo-critical temperature $T_{\chi}\equiv 155\pm2$ MeV \cite{Wupertal2010, WB2010II, HotQCD2012, WB2014, HotQCD2014}.

Since the LQCD \cite{karsch1, karsch2, karsch3, forcrd, varnho, jin, Bazav, jin2} results show large variations due to the strong cut-off and discretization effects,~the exact mapping of the critical line that demarcates the crossover from the  first order transition region in the left corner of the $m_{u,d}-m_{s}$ ($m_{\pi}-m_{K}$) plane of the Columbia plot,~is quite challenging.~For $N_{f}=3$ degenerate quark flavors,~the crossover transition at $\mu=0$ turns first order on the SU(3) symmetric line ($m_{u,d}=m_{s}$ ($m_{\pi}=m_{K}$)) for the critical quark (pion) mass $m_{q} \ (m_{\pi})<m_{q}^{c} (m_{\pi}^{c})$.~The first order region close to the chiral limit is confirmed by the LQCD studies conducted during the 2001 to 2017 and they find the  $m_{\pi}^{c}$ in the range 290-50 MeV \cite{karsch2, karsch3, forcrd, varnho, jin, Bazav, jin2,forcrnd2,Resch} while no direct evidence of the first order transition was found for the 80 MeV $\le m_{\pi} \le$ 140 MeV range in a recent improved LQCD $N_{f}=3$ study \cite{dini21}.~The Ref.\cite {zhang24} finds small $m_{q}^{c}\le 4$ using Möbius domain wall fermions.~Although a small or no first order region is suggested in some improved LQCD studies,~no clear picture has emerged till date as the results still show huge discrepancies owing to the notoriously difficult problem of putting the chiral fermions on the lattice.

The $N_{f}=3$ study in Dyson-Schwinger approach \cite{bernhardt23} and another study in Ref. \cite{kousvos22},~find that the transition in the Columbia plot,~is second order.~Fejos and Hastuda in a recent functional renormalization group (FRG) study using local potential approximation (LPA) \cite{fejos22, fejoHastuda},~found that in contrast to the prediction of $\epsilon$ expansion \cite{rob},~the transition can be of second order for $N_{f}\ge2$ if the $U_A(1)$ symmetry gets restored at the $T_{\chi}$.~But they have cautioned for drawing final conclusions as the LPA completely neglects the wave function renormalization.~The thermal fate of the axial $U_A(1)$ anomaly is still uncertain \cite{lahiri21} as some studies  find that it is relevant at the critical point $T_{\chi}$ \cite{dick15, ding21, kaczmarek21, bazazov12, bha14, kaczmarek23} while others \cite{dini21, brandt16, tomiya16, aoki21, aoki22} claim that it vanishes.~Very recently ,~novel signals were conjectured in Ref.\cite{pisarski24} if the $U_A(1)$  anomaly becomes very weak at the $T_{\chi}$  and  the Ref. \cite{Giacosa} explored the relation,~between the order of the chiral transition and the terms which break the anomalous $U_A(1)$ symmetry.~After constructing the $U_A(1)$ symmetry restoring observables,~their chiral limit behavior have  been  studied in Refs.~\cite{nico13,nicojhep,nico18I,nico18II,solnico}.

The  QCD phase transition investigations performed using the effective models that mimic the symmetries of the QCD Lagrangian,~like the linear sigma model (LSM) \cite{Ortman, Lenagh, Rischke:00, Roder, jakobi, Herpay:05, Herpay:06, Kovacs:2007,Jakovac:2010uy, marko, Fejos},~the quark-meson (QM) model \cite{scav, mocsy, bj, Schaefer:2006ds, SchaPQM2F, Bowman:2008kc,  SchaeferPNP, Schaefer:09,  Schaefer:09wspax, SchaPQM3F, Mao, TiPQM3F,  koch} or the Nambu-Jona-Lasinio \cite{costaA, costaB, fuku08} (NJL) model  and the  non-perturbative FRG technique models \cite{berges, bergesRep, Gies, braunii, pawlanal, fuku11, grahl, mitter,  FejosI, Renke2, brauniii, FejosII, FbRenk, Tripol, Fejos3, Fejos4, fejos5,Herbst},~have a long history of providing important information and valuable insights for the LQCD studies.~One has to circumvent the challenge of finding the light and strange quark or the pion and kaon mass dependence of the model parameters,~if one wants to exploit the QM model as a valuable tool of performing the chiral limit studies.~The most commonly used procedure \cite{Ortman, Lenagh, Schaefer:09,  fuku08, berges, Herbst} known as the fixed-ultraviolet (UV) scheme \cite{Resch} relies on changing the light (strange) explicit chiral symmetry breaking source strengths  $ h_{x}(h_{y}) $ whereas all other parameters are kept same as the ones at the physical point.~In the above scheme,~the spontaneous chiral symmetry breaking (SCSB) disappears in the chiral limit  because the mass parameter $m^2$ becomes positive for the $m_{\sigma} < 800$ MeV \cite{Resch, Schaefer:09}.~Hence the chiral limit can be investigated only for very large scalar $\sigma$ meson masses $m_{\sigma}\ge800$ MeV.~We mention,~here,~that it is not still clear how  the scalar $\sigma$ meson state of the  QM model like Lagrangian,~is related to the physical $\sigma/f_0(500)$ state quoted by  the particle data group (PDG).~The physical $\sigma/f_0(500)$  is a broad resonance state seen  in the $\pi\pi$ scattering.~Its properties and status are discussed in the excellent review by J.R.Pelaez in Ref.~\cite{Pelaez}.~Usually one takes different values of the  scalar $\sigma$ meson mass $m_\sigma$ and compute its effect on the chiral transition.~Resch et. al.\cite{Resch} proposed the ChPT motivated fixed-$f_{\pi}$ scheme to perform the chiral limit study and compute the Columbia plot in the  QM model using FRG framework under the LPA.~In their study,~the initial effective action is heuristically adjusted  to the larger scales ($\Lambda^{\prime} > \Lambda $) for every smaller mass point in the chiral limit path such that the $f_{\pi}$ always remains fixed to its physical value $f_{\pi}$ and hence the SCSB is not lost for the scalar $\sigma$ mass range $m_\sigma=400-600$ MeV.~When the strengths $h_{x}$ and $h_{y}$ are decreased in going to the chiral limit,~the change of scale ($\Lambda^{\prime} > \Lambda $) of the initial effective action in their study accounts for the change in parameters which are always kept  at their physical point value.

The FRG QM Model chiral limit study,~has estimated the impact of quark loop vacuum fluctuations on the extent of crossover,~second and first order transition regions of the Columbia plot,~after switching off the thermal and quantum fluctuations of mesons in the Wetterich equation \cite{wetrich} guided FRG flow of the scale-dependent effective action.~They have emphasized \cite{Resch} that by switching off the meson loops,~they are directly working in the extended mean field approximation (e-MFA) within the FRG which is same as the alternative method of including quark one-loop vacuum term in the QM Model where the vacuum divergences are regularized by the standard dimensional regularization technique and curvature masses of the mesons are used to fix the model parameters as in the Ref. \cite{schafwag12}. The above direct treatment of the fermionic vacuum term in the QM model, has a long history and a large number of such QMVT model studies \cite{Herbst,schafwag12,vac,Gatto,Dima,Anna,lars,guptiw,chatmoh1,vkkr12,TranAnd,chatmoh2,vkkt13,
Weise1,Weyrich,kovacs,Weise2,zacchi1,zacchi2,Rai,Weise3,Gyozo},~have already quantified the strong influence of the fermionic vacuum fluctuations on the QCD phase diagram and the strength of the chiral phase transition occurring at the physical point for experimental $m_{\pi},m_{K},m_{\eta} \text{ and } m_{\eta'}$.~It is important to generate the Columbia plot in the $m_{\pi}-m_{K}$ and quark mass $m_{ud}-m_{s}$ planes of the 2+1 flavor QMVT model by using the $ \mathcal{O}(\frac{1}{f^2})$ accurate infrared regularized U(3) ChPT~\cite{Herpay:05, gasser, borasoyI, borasoyII, Beisert, Becher} scaling relations for the ($m_{\pi},m_{K}$) dependence of $f_{\pi},f_{K} \text{ and } M_{\eta}$ such that the parameter fixing, away from the physical point becomes free from any ambiguity or heuristic adjustment when the $h_{x} \text{ and } h_{y}$
 are reduced in going to the chiral limit.

The  double derivatives of the effective potential at its minimum,~define the curvature masses of the mesons.~Hence the quark one-loop vacuum corrections to meson self energies are included at the zero momentum in the curvature masses as the effective potential is the generator of the n-point functions of the theory at vanishing external momenta.~Noticing the above,~another approach of  treating the quark one-loop vacuum fluctuations consistently,~similar to the work in  Ref.~\cite{laine},~was recently  developed in Refs.~\cite{Adhiand1,Adhiand2,Adhiand3,asmuAnd,RaiTiw,raiti2023} for the two flavor and in Refs.~\cite{vkkr22, skrvkt24} for the 2+1 flavor of the QM model where one finds the on-shell renormalized QM (RQM) model parameters by matching the counter terms in the modified minimal subtraction $\overline{MS}$ scheme with the corresponding counter terms in the on-shell scheme when the running couplings and mass parameter are put into the relation of  the physical (pole) masses of the $\pi,K,\eta, \eta' \text{ and } \sigma$ mesons.~Very recently,~QM model extension with diqaurks,~has been studied in Ref.~\cite{Ander25,Gholami} where the above method of on-shell renormalization of model parameters is applied.~We investigated the chiral limit and generated the $m_{\pi}-m_{K}$ ( $m_{ud}-m_{s}$) planes of the Columbia plot,~in our very recent 2+1 flavor RQM model study~\cite{vkt25} where two input sets of the $ \mathcal{O}(\frac{1}{f^2})$ accurate results from the infrared regularized U(3) ChPT \cite{Herpay:05,gasser, borasoyI, borasoyII, Beisert, Becher} and the standard large $N_c$ U(3) ChPT~\cite{gasser,LeutI, KaisI, herrNPB, herrPLB, Escribano},~were used for obtaining the ($m_{\pi},m_{K}$) dependence of the  $f_{\pi},f_{K} \text{ and } M_{\eta}$ when  one goes away from the physical point.~Though the $\eta'$ meson is treated differently in the two ChPT approaches,~only marginal differences were found in the extent of first,~second order and crossover transition regions of the Columbia plot when the RQM model results from two input sets were compared in Ref.~\cite{vkt25}.~Hence,~use of only the  infrared regularized U(3) ChPT input,~will serve the purpose for computing the QMVT model Columbia plot as proposed in the above paragraph.

The aim of the present work is two-fold : first,~to estimate how  the first, second or crossover chiral transition regions in the $m_{\pi}-m_{K}$($m_{ud}-m_{s}$) and the $\mu-m_{K}$($\mu-m_{s}$) planes of the Columbia plot,~get qualitatively and quantitatively modified when different methods are adopted to treat the quark one-loop vacuum fluctuations  and second,~to disentangle the differences arising due to   different methods of approaching the chiral limit when the quark one-loop vacuum corrections are incorporated by implementing equivalent methods in the QM model.~The 't Hooft coupling $c$ gets renormalized in the RQM model to become significantly large as the condensate dependent part of the $U_A(1)$  anomaly term gets affected when the meson self energies due to quark loops are computed using the pole masses of mesons and parameters are fixed on-shell.~The cubic coupling $c$ does not change in the curvature mass parametrized QMVT model.~This crucial difference of the strength of the $U_A(1)$  anomaly in the QMVT and RQM model will show up in the extent of the first/second order  regions of the Columbia Plot.~Therefore our first goal is to compare the QMVT model Columbia plots ( for the $m_{\sigma}=400 \text{ and }530$ MeV) obtained in the present work  with the corresponding Columbia plots reported for the RQM model \cite{vkt25}.~Since the e-MFA:QM and QMVT model both use curvature  masses of mesons to fix the model parameters,~our second objective of comparing the differences in the first/second order transition regions of the Columbia plots of the QMVT model and  the e-MFA:QM model FRG study for the $m_{\sigma}=530$ MeV \citep{Resch},~will give us the estimate of the differences which arise due to the fact that the $f_{\pi},f_{K}, M_{\eta}$ and parameters,~change towards the chiral limit in the QMVT model (guided by the  ChPT given $(m_{\pi}-m_{K})$ dependence),~whereas the $f_{\pi},~M_{\eta}$ and parameters in the e-MFA:QM-FRG study are kept fixed to their physical point value and the initial effective action is successively adjusted to the higher scales $\Lambda$ as one approaches the chiral limit.~The $m_{\sigma}$ temperature plots will also be compared across the models when $m_{\pi}-m_{K}$ are reduced towards the chiral limit.


The paper is arranged as follows.~The section~\ref{secII} contains $ SU_{L}(3) \times SU_{R}(3) $ QM model with a brief description of QMVT and RQM model respectively in the section~\ref{subsec:QMVT} and~\ref{subsec:RQM}.~The  infrared regularized U(3) ChPT scaling  relations for the $(m_{\pi},m_{K})$ dependence of the $f_{\pi}, f_{K} $ and $M_{\eta}$ are given in the section~\ref{subsec:Chpt}.~The results and discussions are presented in the section~\ref{secIII} where temperature plots of order parameters,~their derivatives and the $\sigma$ mass $m_{\sigma}$,~are compared in the path to chiral limit,~respectively in the section~\ref{subsec:Chlmt} and~\ref{subsec:sigma}.~Various aspects of the QMVT and RQM model Columbia plots for different $m_{\sigma}$ are compared in the  section~\ref{subsec:Colplot}.~The light-strange quark mass $m_{ud}-m_{s}$ and   $\mu-m_{s}$ planes of the QMVT and RQM model Columbia plot for different $m_{\sigma}$,~are compared in the section~\ref{subsec:Colpqm}.~The section~\ref{secIV} contains the summary.
\section{The  Model formulations}
\label{secII}
 The  Lagrangian \cite{Rischke:00,Schaefer:09,TiPQM3F} of the QM model  is
\bqa
\label{lag}
{\cal L_{QM}}&=&\bar{\psi}[i\gamma^\mu D_\mu- g\; T_a\big( \sigma_a 
+ i\gamma_5 \pi_a\big) ] \psi+\cal{L(M)}\;. \\
\nonumber
\label{lagM}
\cal{L(M)}&=&\text{Tr} (\partial_\mu {\cal{M}}^{\dagger}\partial^\mu {\cal{M}}-m^{2}({\cal{M}}^{\dagger}{\cal{M}}))\\ \nonumber
&&-\lambda_1\left[\text{Tr}({\cal{M}}^{\dagger}{\cal{M}})\right]^2-\lambda_2\text{Tr}({\cal{M}}^{\dagger}{\cal{M}})^2\\ 
&&+c[\text{det}{\cal{M}}+\text{det}{\cal{M}}^\dagger]+\text{Tr}\left[H({\cal{M}}+{\cal{M}}^\dagger)\right]\;.
\eqa
The three flavor of quark fields $\psi$ (color $N_c$-plet  Dirac spinor) are coupled by  the Yukawa coupling $g$ to the nine scalar(pseudo-scalar) meson $\xi$ fields $\sigma_a (\pi_a$) of $3\times3$ complex matrix ${\cal{M}}=T_{a} \xi_{a}=T_{a}(\sigma_{a}+i\pi_{a})$.~$T_{a}=\frac{\lambda_{a}}{2}$ where $\lambda_a$ ($a=0,1..8$) are Gell-Mann matrices with $\lambda_0=\sqrt{\frac{2}{3}}{\mathbb I}_{3\times3}$.~The field $\xi$ picks up the non-zero vacuum expectation value $\overline{\xi}$ in the $0$ and $8$ directions.~The condensates $\bar{\sigma_0}$ and $\bar{\sigma_8}$ break the $SU_L(3) \times SU_R(3)$ chiral symmetry spontaneously   while the external fields $H= T_{a} h_{a}$ with $h_0$, $h_8  \neq 0$ break it explicitly.~The change from the singlet octet $(0,8)$  to the non-strange strange basis $(\x,\y)$ gives $\x ( h_{x})= \sqrt{\frac{2}{3}}\bar{\sigma}_0 (h_{0}) +\frac{1}{\sqrt{3}} \bar{\sigma}_8(h_{8})$ and $ \y (h_{y}) = \frac{1}{\sqrt{3}}\bar{\sigma}_0 (h_{0})-\sqrt{\frac{2}{3}}\bar{\sigma}_8 (h_{8})$.~Considering mesons at mean field level  with the thermal and quantum fluctuations of the quarks/anti-quarks,~the grand potential \cite{Schaefer:09,TiPQM3F}, is the sum of the vacuum effective potential $ U(\x,\y) $ and the quark/anti-quark  contribution $\Omega_{q\bar{q}}$ at finite temperature $T$ and quark chemical potential $\mu_{f} (f=u,d,s)$.
\bqa
\label{Grandpxy}
&&\Omega_{\rm MF }(T,\mu)=U(\x,\y)+\Omega_{q\bar{q}} (T,\mu;\x,\y)\;. \\ \nonumber \\ 
&&\Omega_{q\bar{q}}(T,\mu;\x,\y)=\Omega_{q\bar{q}}^{vac}+\Omega_{q\bar{q}}^{T,\mu}. \\ \nonumber
\label{eq:mesop}
&&U(\x,\y)=\frac{m^{2}}{2}\left(\x^{2} +
\y^{2}\right) -h_{x} \x -h_{y} \y
- \frac{c}{2 \sqrt{2}} \x^2 \y \\  
&&+ \frac{\lambda_{1}}{2} \x^{2} \y^{2}+
\frac{1}{8}\left(2 \lambda_{1} +
\lambda_{2}\right)\x^{4} 
+\frac{1}{8}\left(2 \lambda_{1} +
2\lambda_{2}\right) \y^{4}\;. \\ \nonumber \\
\label{vac1}
&&\Omega_{q\bar{q}}^{vac} =- 2 N_c\sum_f  \int \frac{d^3 p}{(2\pi)^3} \ E_q \  \theta( \Lambda_c^2 - \vec{p}^{2})\;.\\ \nonumber \\
\label{vac2}
&&\Omega_{q\bar{q}}^{T,\mu}=- 2 N_c \sum_{f=u,d,s} \int \frac{d^3 p}{(2\pi)^3} T \left[ \ln g_f^{+}+\ln g_f^{-}\right].\; \\ \nonumber \\
\label{GrandQM}
&&\Omega_{\rm QM }(T,\mu,x,y)=U(\x,\y)+\Omega_{q\bar{q}}^{T,\mu}\;. 
\eqa
The $ g^{\pm}_f = \left[1+e^{-E_{f}^{\pm}/T}\right] $ where $E_{f}^{\pm} =E_f \mp \mu_{f}$ and $E_f=\sqrt{p^2 + m{_f}{^2}}$ is the quark/anti-quark energy.~The light (strange) quark mass  $m_{u/d}=\frac{g\x}{2}$ ($m_{s}=\frac{g\y}{\sqrt{2}}$) and $\mu_{u}=\mu_{d}=\mu_{s}=\mu$.~The quark one-loop vacuum term with ultraviolet cut-off $\Lambda_c$ in Eq.~(\ref{vac1}) is dropped for the standard mean field approximation (s-MFA) of the quark meson (QM) model grand effective potential in Eq.~(\ref{GrandQM}).

\subsection{QMVT Model}
\label{subsec:QMVT}
The derivation of the QMVT model effective potential,~is briefly described below.~The curvature masses of the scalar and pseudo-scalar mesons,~are used to fix the parameters in the QMVT model when the quark one-loop vacuum fluctuations in the Eq.~(\ref{vac1}) are included in its effective potential under the the extended mean field approximation (e-MFA).~The vacuum divergence of the Eq.~(\ref{vac1}) is regularized under the minimal subtraction scheme using the dimensional regularization as it was done in the Refs.~\cite{vac,guptiw,vkkr12} for the two flavor and in the Refs.~\cite{schafwag12,chatmoh1,vkkt13} for the three flavor.~After the dimensional regularization of the Eq.~(\ref{vac1}) near three  dimensions, $d=3-2\epsilon$,~one gets :
\begin{equation}
{\hskip -0.1 cm}\Omega_{q\bar{q}}^{\rm vac} =\sum_{f=u,d,s}\frac{N_c\ m_f^4}{16 \pi^2}\left[
  \frac{1}{\epsilon} -\frac{ 
\{-3+2\gamma_E +4\ln(\frac{m_f}{2\sqrt{\pi} \Lambda})\}}{2} \right],
\label{Omega_DR}
\end{equation}
the  $\Lambda$ is  renormalization scale.
\begin{equation}
\delta \mathcal{L} =\sum_{f=u,d,s} \frac{N_c}{16 \pi^2} m_f^4 \left[ \frac{1}{\epsilon} - \frac{1}{2}
 \left\{ -3 + 2 \gamma_E - 4 \ln (2\sqrt{\pi})\right\} \right], 
\label{counter}
\end{equation}
Adding the counter term  $\delta \mathcal{L}$ to the QM model Lagrangian,~one gets the  quark one-loop vacuum term as: 
\begin{equation}
\Omega_{q\bar{q}}^{\rm vac} =  -\sum_{f=u,d,s} \frac{N_c}{8 \pi^2} m_f^4  \ln\left(\frac{m_f}{\Lambda}\right)\;.
\label{Omega_reg} 
\end{equation}
The vacuum grand potential becomes the  scale $\Lambda$ dependent after including the above vacuum correction  as :
\bqa
\Omega^{\Lambda} (x,y) =U(x,y)+\Omega_{q\bar{q}}^{\rm vac}.
\label{Omeg_rel}
\eqa
The $x$ and $y$ dependent curvature masses of the mesons determine the six unknown parameters $m^2$, $\lambda_1, \lambda_2$, $h_x, h_y$ and $c$ in the meson 
potential U($x,y$).~The details of the parameter fixing can be found in the Ref.~\cite{vkkt13}.~The parameter $\lambda_2$ is determined as $\lambda_2=\lambda_{2s}+n+\lambda_{2+}+\lambda_{2\Lambda}$ where $\lambda_{2s}$ is the same old $\lambda_2$ parameter of the QM/PQM model in the Ref.~\cite{Rischke:00,Schaefer:09,TiPQM3F},~$n=\frac{N_cg^4}{32\pi^2}$, $\lambda_{2+}=\frac{n{f_{\pi}}^2}{f_{K} \left(f_{K}-f_{\pi}\right)}\ln\{\frac{2 f_K-f_{\pi}}{f_{\pi}}\}$ and $\lambda_{2\Lambda}= 4n\ln\{ \frac{g\left( 2f_K-f_{\pi}\right)}{2 \Lambda}\}$.~The renormalization scale $\Lambda$ dependent part $\lambda_{2\Lambda}$ in the above is generated by the logarithmic $\Lambda$ dependence in the term $\Omega_{q\bar{q}}^{\rm vac}$.~After substituting this value of the $\lambda_2$  in the expression of U($x,y$) in the Eq.~(\ref{eq:mesop}) and writing explicitly all the terms under the summation in the expression of $\Omega_{q\bar{q}}^{\rm vac}$,~the Eq.~(\ref{Omeg_rel}) becomes :
\bqa
\label{eq:mesVeffLamb}
\nonumber
{\hskip -1.0 cm}\Omega^{\Lambda} (x,y)&=&\frac{m^{2}}{2} \left( x^{2} +
y^{2}\right) -h_{x} x-h_{y} y
-\frac{c}{2 \sqrt{2}} x^2 y  \\ \nonumber
&& +\frac{\lambda_{1}}{4}\left(x^{4} +y^{4}+2 x^{2} y^{2}\right) \\ \nonumber
&&+\frac{\left(\lambda_{2\text{v}}+n+\lambda_{2\Lambda} \right)}{8} \left(x^{4} + 2 y^{4} \right)
\\  
&&-\frac{n x^4}{2}\ln\left(\frac{g \ x}{2\Lambda}\right)-\ n y^4\ln\left(\frac{g \ y}{\sqrt{2}\Lambda}\right)\;.   
\eqa 
When one does the rearrangement of the terms,~one finds that the scale dependence of all the terms in $\Omega_{q\bar{q}}^{\rm vac}$ gets completely canceled by the logarithmic $\Lambda$ dependence of the $\lambda_{2}$ contained in $\lambda_{2\Lambda}$.~The  cancellation of the scale $\Lambda$ independence is obtained numerically in the Ref.\cite{schafwag12}.~The scale independent analytical expression of the vacuum effective potential as derived in Ref.~\cite{vkkt13} is written as :
\bqa
\nonumber
 \Omega(x,y) &=& \frac{m^{2}}{2}\left(x^{2} +
  y^{2}\right) -h_{x} x -h_{y} y
 - \frac{c}{2 \sqrt{2}} x^2 y \\ \nonumber 
 && +\frac{\lambda_{1}}{4} \left( x^{4} + y^{4} +2 x^{2} y^{2} \right)
    + \frac{ \left( \lambda_{2\text{v}}+n \right) }{8}\left( x^{4} + 2 y^{4} \right)  \\ \nonumber
 && -\frac{n x^4}{2} \ln \left( \frac{x}{\left(2f_K-f_{\pi} \right) } \right) \\
 && - n y^4 \ln \left( \frac{\sqrt 2\ y}{\left(2f_K-f_{\pi}\right)}\right)\;.
\label{eq:mesVop}
\eqa
The $\lambda_{2\text{v}}=\lambda_{2s}+\lambda_{2+}$.~Note that the quark one-loop vacuum corrections,~in the QMVT model parameter fixing scheme,~modify the parameters  $m^2$, $\lambda_1$ and $\lambda_2$ and  the unaffected parameters $h_x$, $h_y $ and $c$ remain the same as in the QM model.~The QMVT model thermodynamic grand potential is written as :
\bqa
\Omega_{\rm QMVT}(T,\mu;x,y) & =&  \Omega(x,y) + \Omega_{q\bar{q}}^{\rm T ,\mu}\;.
\label{OmegaMFQMVT}
\eqa
The global minimum of the grand potential for a given  $\mu \text{ and } T$ gives the $x$ and $y$.
\bqa 
\label{eq:gapeq}
\nonumber
\frac{ \partial \Omega_{\rm QMVT} (T,\mu;x,y)}{\partial x}|_{x, y}  &=& \frac{ \partial \Omega_{\rm QMVT} (T,\mu;x,y)}{\partial y}|_{x,y} = 0 \\ .
\eqa 
 
Note that the curvature mass scheme of parameter fixing,~does not consider the dressing of the meson propagators.~Therefore the $f_\pi $ and $f_{K} $ do not get renormalized.~The parameters of the effective potential,~are modified by the quark one-loop vacuum correction in such a way that the the non-strange and strange direction stationarity conditions at the $T=0$,~give the same  $h_x$ and $h_y$ that one gets in the QM model.~The modified $\pi \text{ and } K$ curvature masses turn out to be same as their pole masses as one can find in the Appendix (B) of Ref.~\cite{vkkt13}.~The minimum of the vacuum effective potential does not shift and it remains at the $ \overline{x}=f_\pi $ and $ \overline{y}=\frac{(2f_K-f_\pi)}{\sqrt{2}} $.

\subsection{RQM model}
\label{subsec:RQM}

Note that the curvature meson mass calculation,~involves the determination of the meson self-energies at zero momentum because the effective potential is the generator of the n-point functions of the theory at vanishing external momenta \cite {laine, Adhiand1, Adhiand2, Adhiand3, asmuAnd, RaiTiw, raiti2023}.~One should also take into account that the pole definition of the meson mass is the physical and gauge invariant \cite{Kobes, Rebhan} one.~In the light of the above considerations,~the consistent e-MFA RQM model effective potential was  calculated in our very recent works \cite{vkkr22,skrvkt24}  after relating the counter-terms in the $\overline{\text{MS}}$ scheme to those in the on-shell (OS) scheme \cite{Adhiand1, Adhiand2, Adhiand3, asmuAnd, RaiTiw, raiti2023}.~In order to find the relations between the renormalized parameters of both the schemes,~the $\overline{\text{MS}}$ running  couplings and mass parameter are put into the relation of the physical quantities (pole masses  $m_{\pi}, m_{K}, m_{\eta},m_{\eta^{\prime}} \ \text{and} \ m_{\sigma}$,~the pion and kaon decay  constants $f_{\pi}$ and $f_{K}$).~These relations are used as input when the effective potential is calculated using the modified minimal subtraction procedure.~After the cancellation of the  $ 1 / \epsilon$ divergences,~the vacuum effective potential 
$ \Omega_{vac}=U(x_{\ms},y_{\ms})+\Omega^{q,vac}_{\ms}+\delta U(x_{\ms},y_{\ms})$ in the $\overline{\text{MS}}$ scheme has been rewritten in Ref. \cite{vkkr22} in terms of the scale $\Lambda$ independent constituent quark mass parameters $\Delta_{x}=\frac{g_{\ms} \ x_{\ms}}{2}$ and $\Delta_{y}=\frac{g_{\ms} \ y_{\ms}}{\sqrt{2}}$ as the following.
\begin{align}
\label{OmegDelxy}
\nonumber 
&\Omega_{\rm vac}(\Delta_{x},\Delta_{y})=\frac{m^2_0}{g^2_0}(2\Delta_{x}^2+\Delta_{y}^2)-2\frac{h_{x0}}{g_0}\Delta_{x}-\sqrt{2}\frac{h_{y0}}{g_0}\Delta_{y}\\ \nonumber 
&-2\frac{c_{0}}{g^3_{0}}\Delta_{x}^2 \ \Delta_{y}+4\frac{\lambda_{10}}{g^4_{0}}\Delta_{x}^2 \ \Delta_{y}^2+2\frac{(2 \lambda_{10}+\lambda_{20})}{g^4_{0}}  \Delta_{x}^{4} \ \\ \nonumber 
&+\frac{(\lambda_{10}+\lambda_{20})}{g^4_{0}}  \Delta_{y}^{4}+
\frac{2N_c\Delta_{x}^4}{(4\pi)^2}\left[\frac{3}{2}+\ln\left(\frac{\Lambda^2}{m_{u}^2}\right)+\ln\left(\frac{m_{u}^2}{\Delta_{x}^2}\right)\right] \\ 
&+\frac{N_c\Delta_{y}^4}{(4\pi)^2}\left[\frac{3}{2}+\ln\left(\frac{\Lambda^2}{m_{u}^2}\right)+\ln\left(\frac{m_{u}^2}{\Delta_{y}^2}\right)\right]\;.  
\end{align}
The condition that the minimum of the effective potential of the RQM model does not shift from that of the QM model fixes the scale $\Lambda_0$ \cite{vkkr22,skrvkt24} as :
\bqa
\label{Sclcond}
&\ln\left(\frac{\Lambda^2_0}{m_u^2}\right)+\mathcal{C}(m^2_\pi)+m^2_\pi \mathcal{C}^{\prime}(m^2_\pi)=0\;. 
\eqa
The terms $\mathcal{C}(m^2_\pi) $ and $ \mathcal{C}^{\prime}(m^2_\pi) $ and the derivations for the renormalized parameters  $m^{2}_{0}=(m^{2}+m^2_{\text{\tiny{FIN}}})$,~$h_{x0}=(h_{x}+h_{x\text{\tiny{FIN}}})$,~$h_{y0}=(h_{y}+h_{y\text{\tiny{FIN}}}) $,~$\lambda_{10}=(\lambda_{1}+\lambda_{1\text{\tiny{FIN}}})$,~$\lambda_{20}=(\lambda_{2} +\lambda_{2\text{\tiny{FIN}}})$ and $c_{0}=(c+c_{\text{\tiny{FINTOT}}})$ are given in detail in Refs. \cite{vkkr22,skrvkt24}.~The $m^2_{\text{\tiny{FIN}}}$,
$ h_{x\text{\tiny{FIN}}} $, $ h_{y\text{\tiny{FIN}}} $, $\lambda_{1\text{\tiny{FIN}}} $, $ \lambda_{2\text{\tiny{FIN}}}$ and 
$c_{\text{\tiny{FINTOT}}}$ are the finite on-shell corrections in the parameters at the scale $\Lambda_0$.~The experimental values of the pseudo-scalar meson masses $m_{\pi}$, $m_{K}$, $m_{\eta}$, $m_{\eta^{\prime}} $, the scalar $\sigma$ mass $m_{\sigma}$ and the $f_{\pi}$, $f_K$ as input determine  the tree level QM model quartic couplings  $\lambda_1$, $\lambda_2$,~mass parameter  $m^2$,~$h_x$, $h_y$ and the coefficient $c$ of the 't Hooft determinant term for the $U_A(1)$ axial anomaly \cite{Rischke:00,Schaefer:09}.

~Although the $f_{\pi}$, $f_{K}$ and $g$ get renormalized due to the dressing of the meson propagator in the on-shell scheme,~they do not change as the $g_{\ms}=g_{ren}=g_{0}=g$, $x_{\ms}=x$, $y_{\ms}=y$ at $\Lambda_0$.~Further  $x_{\ms}=f_{\pi,ren}=f_\pi$ and $y_{\ms}=\frac{2f_{K,ren}-f_{\pi,ren}}{\sqrt{2}}$= $\frac{2f_K-f_\pi}{\sqrt{2}}$ at the minimum.~Using $\Delta_{x}=\frac{g \ x}{2}$ and $\Delta_{y}=\frac{g \ y}{\sqrt{2}}$,~the vacuum effective potential in the Eq.~(\ref{OmegDelxy}) is  written in terms of the $x$ and $y$ as :
\begin{align}
\label{vacRQM}
\nonumber
&\Omega_{vac}^{\rm RQM}(\x,\y)=\frac{(m^{2}+m^2_{\text{\tiny{FIN}}})}{2} \ \left(\x^{2} +
\y^{2}\right)-(h_{x}+h_{x\text{\tiny{FIN}}}) \ \x 
\\ \nonumber 
&-(h_{y}+h_{y\text{\tiny{FIN}}}) \y -\frac{(c+c_{\text{\tiny{FINTOT}}})}{2 \sqrt{2}} \x^2 \y + \frac{(\lambda_{1}+\lambda_{1\text{\tiny{FIN}}})}{2} \x^{2} \y^{2}
\\ \nonumber
&+\frac{\left\{2(\lambda_{1} +\lambda_{1\text{\tiny{FIN}}})+
( \lambda_{2} +\lambda_{2\text{\tiny{FIN}}})\right\}\x^{4}}{8}+\left( \lambda_{1} +\lambda_{1\text{\tiny{FIN}}}+\lambda_{2}
 \right.  \\  \nonumber
& \left. +\lambda_{2\text{\tiny{FIN}}}\right)\frac{ \y^{4}}{4} +\frac{N_c g^4 (\x^4+2\y^4)}{8(4\pi)^2} \left[\frac{3}{2}-\mathcal{C}(m^2_\pi)-m^2_\pi \mathcal{C}^{\prime}(m^2_\pi)\right]\ \\
&-\frac{N_c g^4 }{8(4\pi)^2} \left[\x^4 \ln\left(\frac{\x^2}{f_{\pi}^2}\right)+2\y^4 \ln\left(\frac{2 \  \y^2}{f_{\pi}^2}\right) \right]. \\ \nonumber  \\ 
\label{grandRQM}
&\Omega_{\rm RQM }(T,\mu,x,y)=\Omega_{vac}^{\rm RQM}(\x,\y)
+\Omega_{q\bar{q}}^{T,\mu}\;.
\end{align}
The search of the grand potential minima  $\frac{\partial \Omega_{\rm RQM}}{\partial \x}= \frac{\partial \Omega_{\rm RQM}}{\partial \y}=0$ for the  Eq.~(\ref{grandRQM}) gives the  $T$ and  $\mu$ dependence of the $ \x$ and $ \y$.~The curvature masses of mesons are different from their pole masses in the RQM model due to the on-shell renormalization of the parameters \cite{BubaCar,fix1}.~The derivation of the RQM model  curvature mass expressions for the mesons are presented in the section (IIA) of the Ref.\citep{vkt25}.~Using the Eq.~(\ref{vacRQM}) to evaluate
the equations of motion $\frac{\partial \Omega_{vac}^{\rm RQM}}{\partial \x}=0= \frac{\partial \Omega_{vac}^{\rm RQM}}{\partial \y}$,~one gets the renormalized explicit chiral symmetry breaking strengths $h_{x0}$ and $h_{y0}$ as the following.
\bqa
\label{hx0}
&h_{x0}=m_{\pi,c}^2 \ f_{\pi}. \\
\label{hy0}
&h_{y0}=\biggl(\sqrt{2} \ f_K \ m^2_{K,c}-\frac{f_{\pi}}{\sqrt{2}} \ m^2_{\pi,c}\biggr).
\eqa 
The pion and kaon curvature masses $m_{\pi,c}$ and $m_{K,c}$ as derived in the Ref. \cite{vkkr22} have the following expressions.
\begin{align}
\label{mpicr}
&m_{\pi,c}^2=m_{\pi}^2\biggl\{ 1-\frac{N_{c}g^2}{4\pi^2} \ m_{\pi}^2 \ \mathcal{C}^{\prime}(m^2_\pi,m_u)  \biggr\}. \\ \nonumber
\label{kcr}
&m_{K,c}^2= m_{K}^2 \biggl[ 1-\frac{N_{c}g^2}{4\pi^2} \biggl\{  \mathcal{C}(m^2_\pi,m_u)+m^2_\pi \ \mathcal{C}^{\prime}(m^2_\pi,m_u)  \\ \nonumber
&\hspace*{.8cm}-\biggl(1-\frac{(m_s-m_u)^2}{m_K^2}\biggr)\ \mathcal{C}(m^2_K,m_u,m_s)+ \\  &\hspace*{.8cm}\biggl(1-\frac{f_\pi}{f_K}\biggr) \ \biggl(\frac{m^2_u-m_s^2+2m_s^2\ln(\frac{m_s}{m_u})}{m^2_K}\biggr)\biggr\}\biggr]. 
\end{align}
The $\pi \text{ and } K$ curvature masses  $m_{\pi;c}= 135.95$ MeV and $m_{K;c}= 467.99$ MeV are, respectively, smaller by 2.05 MeV and 28.01 MeV from their corresponding pole masses  $m_{\pi}= 138$ MeV and $m_{K}= 496$  MeV \citep{vkkr22,skrvkt24}.

\subsection{The ChPT scaling of  $\bf f_{\pi},f_{K} \ \text{and} \ M_{\eta}^2=m_{\eta}^2+m_{\eta^{\prime}}^2$}
\label{subsec:Chpt}

The degrees of freedom in the perturbative QCD regime are $u,d$ and $s$ quarks while the $2+1$ flavor QM model,~deals with the low energy scalar and pseudo-scalar mesons.~Hence,~the current quark mass ($m_{ud}-m_{s}$ plane) sensitivity of the chiral transition in  the chiral limit,~can be studied by computing the Columbia plot in the  equivalent description of the $m_{\pi}-m_{K}$ plane.~The model parameters as well as the light and strange explicit chiral symmetry breaking source strengths $h_{x}=f_{\pi} m^2_{\pi}$,~$h_{y}=\sqrt{2} f_{K} m^2_{K}-f_{\pi} m^2_{\pi}/\sqrt{2}$ can be determined only at the physical point where the experimental information compatible with the underlying QCD theory is available.~Since,~a priori,~it is not clear how the parameters change as the system is tuned away from the physical point,~different strategies are used to fix the model parameters in going to chiral limit.~The often used method \cite{Ortman, Lenagh, Schaefer:09, fuku08, berges,Herbst},~termed as the fixed-ultraviolet (UV) scheme \cite{Resch},~relies on the assumption that the change in the current masses of quarks in the QCD regime can be directly mapped onto a change of  symmetry breaking source strengths.~Hence the $h_{x}\text{ and }h_{y}$ are varied while  all other parameters are kept fixed to their physical point value such that the initial effective action does not change in the UV \cite{Resch}.~Since the mass parameter $m^2$ becomes positive in the scalar $\sigma$ mass range  $m_{\sigma}=400-800$ MeV,~the fixed-UV scheme does not work due to the loss of the spontaneous chiral symmetry breaking (SCSB) in the chiral limit \cite{Schaefer:09,Resch}.~In the chiral limit studies of Refs. \cite{Schaefer:09},~this problem was circumvented by choosing quite a large $m_{\sigma}\ge800$ MeV.

Resch et.al. \cite{Resch} in a nonperturbative FRG study first of its kind,~proposed a ChPT  motivated fixed-$f_{\pi}$ scheme where they heuristically adjust the initial effective action to the larger scales ($\Lambda^{\prime} > \Lambda $) while traversing 
the chiral limit path of smaller masses in the Columbia plot such that the infra-red (IR) pion decay constant  $f_{\pi}$ always retains it fixed value at physical point.~The problem of fixed-UV scheme gets resolved by construction,~in the fixed-$f_{\pi}$ scheme,~as the condensates being intimately related to the fixed $f_{\pi}$ do not drop to zero and the SCSB is not lost.~In the fixed-$f_{\pi}$ scheme,~the change of scale ($\Lambda^{\prime} > \Lambda $) is equivalent to the change of parameters (which retain the same value at the $\Lambda $ in the fixed UV-scheme) when the $h_{x} \text{ and } h_{y}$  are decreased away from the physical point.~While being physically reasonable,~the fixed $f_{\pi}$ scheme,~suffers from the problem of a heuristic determination of the initial effective action \cite{Resch}.



%

The QCD chiral symmetry becomes $ U_{L}(3) \times U_{R}(3) $ in the large $N_{c}$ limit
as the quark loops causing the $U_{A}(1)$ anomaly are suppressed.~Constructing effective Lagrangian with the $\eta'$  as ninth Goldstone boson,~a systematic expansion of the Green functions of the U(3) ChPT in powers of  momenta, quark masses and the $ \mathcal{O}(p^2)$ parameter $\frac{1}{N_{c}}$,~was achieved in Refs.\cite{gasser,LeutI, KaisI, herrNPB, herrPLB, Escribano}.~The Refs.~\cite{borasoyI, borasoyII, Beisert} argued that the effective Lagrangian can be constructed and treating  perturbatively,~the $\eta'$ can be systematically included in the  ChPT without using the $\frac{1}{N_{c}}$ counting rules as the additional energy scale of the $m_{\eta'} \sim 1$ GeV is proportional to the  $\frac{1}{N_{c}}$.~This alternative treatment of massive fields was  firmly established in Ref. \cite{Becher} where the  Lorentz and chiral invariance are kept at all stages \cite{borasoyI, borasoyII} when the loop integrals are evaluated using a modified regularization scheme,~the so-called infrared regularization.~The authors in  Ref. \cite{borasoyII} point out that the $\eta'$ loops do not contribute (at the fourth order Lagrangian) in infrared regularization and the $\eta'$ can be treated as a background field while the contribution of Goldstone boson ($\pi,K,\eta$) loops  get evaluated.~The above approach is different from the
large $N_c$ standard U(3) ChPT works in Refs. \citep{herrNPB, herrPLB, Escribano} where  the $\eta'$  mesons are included in the loops as the original Goldstone Bosons.

~In our very recent 2+1 flavor RQM model work~\cite{vkt25},~we investigated the chiral limit and generated the $m_{\pi}-m_{K}$ ( $m_{ud}-m_{s}$) planes of the Columbia plot where two input sets of the $ \mathcal{O}(\frac{1}{f^2})$ accurate results from the infrared regularized U(3) ChPT termed as the input set $\text{M}_{\eta}$-I in Ref.\cite{vkt25} and the standard large $N_c$ U(3) ChPT termed as the input set $\text{M}_{\eta}$-II in Ref.\cite{vkt25},~were used for obtaining the ($m_{\pi},m_{K}$) dependence of the  $f_{\pi},f_{K} \text{ and } M_{\eta}$ when  one moves away from the physical point.~Though the $\eta'$ meson is treated differently in the two ChPT approaches,~only marginal differences were found in the extent of first,~second order and crossover transition regions of the Columbia plot when the RQM model results from two input sets were compared in Ref.~\cite{vkt25}.~Therefore in the present QMVT model study,~the chiral limit and  the $m_{\pi}-m_{K}$ ($m_{ud}-m_{s}$) planes of the Columbia plot,~will be investigated using only the  infrared regularized U(3) ChPT given ($m_{\pi},m_{K}$) dependent expressions of the  $f_{\pi},f_{K} \text{ and } M_{\eta}$ which are briefly reproduced below.

Originally proposed for the linear sigma model studied in the context of  optimized perturbation theory,~this ChPT based parameter fixing method in Ref.\cite{Herpay:05},~ uses the infrared regularized U(3) ChPT expressions for $M_{\eta}^2$.~The functional forms $f_{\pi} (m_\pi, m_K )    \text{ and } f_{K}(m_\pi, m_K )$ are needed for which it is sufficient to use the $ SU_{L}(3) \times SU_{R}(3) $ ChPT framework in this approach.~The $ \mathcal{O}(\frac{1}{f^2})$ accurate expressions of $m_\pi^2, m_K^2, f_{\pi}, f_{K}$  in terms of the eight parameters $f, A, q, M_{0}, L_{4}, L_{5}, L_{6}, L_{8}$ given by the one-loop ChPT calculation \cite{gasser} are the following. 
\bqa
\label{mpi2}
\nonumber
m_{\pi}^2&=&2A\left[1+\frac{1}{f^2}\biggl\{ \mu_{\pi}-\frac{\mu_{\eta}}{3} +16A(2 L_{8}-L_{5}) \right. \\ 
&&\left.+16A(2+q)(2L_{6}-L_{4})\biggr\}\right]\;. \\ \nonumber
\label{mk2}
m_{K}^2&=&A(1+q)\left[1+\frac{1}{f^2}\biggl\{\frac{2}{3} \mu_{\eta} +8A(1+q)(2 L_{8}-L_{5}) \right. \\ 
&&\left.+16A(2+q)(2L_{6}-L_{4})\biggr\}\right]\;. 
\eqa
\bqa
\label{fpiol}
f_{\pi}&=&f+\frac{1}{f} \left[-2\mu_{\pi}-\mu_{K}+8AL_{5}+8A(2+q) L_{4}\right]. \\ \nonumber
\label{fkol}
f_{K}&=&f \left[1+\frac{1}{f^2}\biggl\{-\frac{3}{4}(\mu_{\pi}+\mu_{\eta}+2\mu_{K}) \right. \\  
&&\left.\ \ \ +8A(2+q) L_{4}  +4A(1+q) L_{5} \biggr\} \right]\;. 
\eqa
In the above,~the chiral logarithms $\mu_{\text{\tiny{PS}}}=\frac{m_{\text{\tiny{PS}}}^2}{32 \pi^2} \ln(\frac{m_{\text{\tiny{PS}}}^2}{M_{0}^2})$ are at the scale $M_{0}$ and for the $ m_{\text{\tiny{PS}}}^2$,~one takes the leading order squared mass of the corresponding meson of the pseudo-scalar octet.~Recall that the chiral constants $L_i$ do not vary with the pseudo-scalar masses.~The parameters  $q=2 m_s /(m_u+m_d)=m_s/m_{ud}$ and $A=B \ (m_u+m_d)/2=B \ m_{ud}$ where the constant $B$ is determined by the quark condensate $<\bar{u} u> $ in the chiral limit,~govern the quark mass and quark condensate dependencies of the above one-loop  ChPT expressions of the $m_\pi^2, m_K^2, f_{\pi} \ \text{and} \ f_{K}$.~The $q$ and $A$ can be expressed below in terms of the masses $m_\pi, m_K, m_{\eta}$  and chiral constants $L_i$ after inverting the Eqs.~(\ref{mpi2}) and (\ref{mk2}) to order $\mathcal{O}(\frac{1}{f^2})$ accuracy.
\bqa
\label{Aqpim}
\nonumber
A&=&\frac{m_{\pi}^2}{2}\left[1-\frac{1}{f^2}\biggl\{ \mu_{\pi}-\frac{\mu_{\eta}}{3} +8m_{\pi}^2(2 L_{8}-L_{5}) \right. \\ 
&&\left.+8(2m_{K}^2+m_{\pi}^2)(2L_{6}-L_{4})\biggr\}\right]\;. \\ \nonumber
\label{qmr}
q+1&=&\frac{2m_{K}^2}{m_{\pi}^2}\left[1-\frac{1}{f^2}\biggl\{\mu_{\eta} -\mu_{\pi} \right. \\ 
&&\left.+8(m_K^2-m_{\pi}^2)(2 L_{8}-L_{5})\biggr\}\right]\;.
\eqa
The leading order relations of the above two equations are sufficient to find the following $\mathcal{O}(\frac{1}{f^2})$ accurate $m_{\pi},  m_{K}$-dependence of the $f_{\pi}, f_{K}$ from Eqs.~(\ref{fpiol}) and (\ref{fkol}). 
\bqa
\label{fpi}
f_{\pi}&=&f-\frac{1}{f} \biggl[2\mu_{\pi}+\mu_{K}-4m_{\pi}^2(L_{4}+L_{5})-8m_{K}^2 L_{4}\biggr].  \hspace {.6 cm} \\  \nonumber
\label{fk}
f_{K}&=&f \left[1-\frac{1}{f^2}\biggl\{\frac{3}{4}(\mu_{\pi}+\mu_{\eta}+2\mu_{K})-4m_{\pi}^2 L_{4} \right. \\  
&&\left.\ \ \  -4m_{K}^2(2L_{4}+L_{5}) \biggr\} \right]\;.  
\eqa
 With the input $f_\pi$=93 MeV, $f_K$=113 MeV,  $m_\pi$=138 MeV, $m_K$=495.6 MeV,$m_\eta$=547.8 MeV and $M_{0}=4 \pi f_{\pi}\equiv 1168$ MeV, $f$=88 MeV, one gets the constants $L_4$ and $L_5$ at the physical point as :
\bqa
L_4=-0.7033 \times 10^{-3} \ ; \ \ \ \  L_5=0.3708 \times 10^{-3}.
\eqa 
The chiral constants $L_6$ and $L_8$ are controlled by the values of $A$ and $q$ taken at the physical point.~The ratio of the strange to average light quark mass has the value $q=24.9$ similar to the Ref.\cite{Herpay:05} as it is close to the lattice determination and compatible with the range $ 20 \le q \le 34$ indicated by the PDG listing \cite{Eidel}.~Choosing leading order ChPT value of $A$ in the physical point $A=A^{(0)}$ and then using the phenomenological values of $m_{\pi}^2, m_{K}^2$ with the Gell-Mann-Okubo formula for $m_{\eta}^2$ in the $\mathcal{O}(\frac{1}{f^2})$ accurate expressions of $A$ and $q$,~one gets :
\bqa
L_6=-0.3915 \times 10^{-3} \ ; \ \ \ \  L_8=0.511 \times 10^{-3}.
\eqa
The values of $M_{0}$,~$f$ and $L_i$  can be used further for the continuation of $A$ and $q$ from the physical point to an arbitrary point in the $m_{\pi}-m_K$ plane.

Recall that the mixing of the singlet $\eta_{0}$ and octet $\eta_{8}$ states,~give the physical states of $\eta$ and $\eta^{\prime}$.~In addition to the $m_{\pi}^2$ and $m_{K}^2$ values,~the numerical value of the $m_{\eta,00}^2+m_{\eta,88}^2$ is needed in the input to  find the quartic coupling $\lambda_{2}$ of the sigma/QM model \cite{Lenagh,Schaefer:09,Mao,TiPQM3F}.~The 't Hooft cubic coupling $c$ is calculated easily if $\lambda_{2}$ is known.~The experimental masses $m_{\eta}$ and $m_{\eta^{\prime}}$ are known only at the physical point.~Since the total sum $m_{\eta}^2+m_{\eta^{\prime}}^2=m_{\eta,00}^2+m_{\eta,88}^2=M_{\eta}^2$ is relevant and needed for finding the $\lambda_{2}$ as one moves away from the physical point,~even if we do not know the $m_{\eta}^2 \ (m_{\pi},m_{K}) $ and $m_{\eta^{\prime}}^2 \ (m_{\pi},m_{K}) $ individually,~the knowledge of the sum $M_{\eta}^2  \  (m_{\pi},m_{K})$ as input is necessary.

The functional dependence $M_{\eta}^2  \  (m_{\pi},m_{K})$,~is obtained by the application of the infrared regularized $ U_{L}(3) \times U_{R}(3) $ ChPT where similar  steps are implemented as in the  previous description but,~here,~the mass mixing in the $\eta_{0}, \eta_{8}$ sector makes it a little involved.~The collections of relevant formulas spread over several papers \cite{herrNPB,borasoyI,borasoyII} are given in Ref. \cite{Herpay:05}.~Taking the mixing angle $\theta_{\eta}=-20^{0}$ and experimental values for masses $m_{\eta}$,~$m_{\eta^{\prime}}$,~one gets the physical point values of the $m_{\eta_{00}}^2, m_{\eta_{88}}^2, m_{\eta_{08}}^2 $.~Putting  their respective numerical values at the physical point into the corresponding ChPT expressions of the  $m_{\eta_{00}}^2, m_{\eta_{88}}^2, m_{\eta_{08}}^2 $,~the three relations restrict  four chiral constants $L_{7}, v_{0}^{(2)}, v_{2}^{(2)}, v_{3}^{(1)} $ appearing in those expressions.~Note that the constant $ v_{0}^{(2)}$ takes into account the $U_{A}(1)$ anomaly contribution to the $\eta'$ mass which is largely determined  by the topological  configurations of the gluons.~Hence it would be insensitive against the quark mass variations.~The choice of the large $N_c$ relation for the $ v_{0}^{(2)}=-29.3 f^2$ in the Ref.~\cite{Herpay:05},~give the remaining three chiral constants as :
\bqa
\hspace{-.2 cm }L_7=-0.2272 \times 10^{-3} ; \ v_{3}^{(1)}=0.095 ; \ v_{2}^{(2)}=-0.1382. 
\eqa
The sum of Eqs. (B10) and (B11) in Ref. \cite{Herpay:05} gives the following ($m_{\pi},m_{K}$) dependence of the $M_{\eta}^2 $.	 
\bqa
\label{Meta}
\nonumber
M_{\eta}^2&=&2m_{K}^2-3v_{0}^{(2)}+2(2m_{K}^2+m_{\pi}^2)\
\{ 3v_{2}^{(2)}-v_{3}^{(1)} \}+  \\ \nonumber 
&&\hspace {-.4 cm}\frac{1}{f^2} \biggl[8v_{0}^{(2)}(2m_{K}^2+m_{\pi}^2)(3L_{4}+L_{5})+m_{\pi}^2(\mu_{\eta}-3\mu_{\pi})  \\ \nonumber
&&\hspace {-.4 cm}-4m_{K}^2 \mu_{\eta}+\frac{16}{3}\biggl\{(6L_{8}-3L_{5}+8L_{7})(m_{\pi}^2-m_{K}^2)^2+\\
&&\hspace {-.4 cm}2L_{6}(m_{\pi}^4-2m_{K}^4+m_{\pi}^2m_{K}^2)+L_{7}(m_{\pi}^2+2m_{K}^2)^2\biggr\} \biggr]\;.     \\ \nonumber
\eqa

The $\eta$ mass occurs at the leading order in chiral logarithm i.e.  $(m_{\eta}^{(0)})^{2}=(4 m_{K}^2-m_{\pi}^2)/3$,~hence the functions $f_{\pi}(m_{\pi},m_{K})$, $f_{K} (m_{\pi},m_{K})$ and  $M_{\eta}^2 (m_{\pi},m_{K})$ are applicable only when $4 m_{K}^2 >m_{\pi}^2 $ .~The Eqs.(\ref{fpi}) and (\ref{fk}) together with the Eqs.(\ref{Meta}),~enable the calculation of the QM/QMVT/RQM model couplings $\lambda_{2}, c $,$m^{2}$ and $\lambda_{1}$ in the $m_{\pi}-m_{K}$ plane.

The reduction of $m_{\pi}$ and $m_{K}$ gives the reduced chiral symmetry breaking strengths in  the chiral limit path.~The  $m_{\pi}-m_{K}$ plane results can be mapped onto the light-strange quark mass $m_{us}-m_{s}$ plane using the Eqs.~(\ref{Aqpim}) and (\ref{qmr}) whose multiplication gives  $\mathcal{O}(1/f^2)$ expression of  $A(q+1)=B(m_{s}+m_{ud})$ in terms of the $m_\pi,m_K$ and chiral constants.~The $m_{\pi}=138$ MeV determines the constant $B$ from the Eq.~(\ref{Aqpim}) as $A=B \ m_{ud}$ by taking the  $m_{ud}=4$ MeV at the physical point.~Once the constant $B$ is known,~one can find quark masses  away from the physical point and  the Columbia plot in the $m_{ud}-m_{s}$ plane can be  generated easily. 

\section{Results and Discussions}
\label{secIII}

Several QMVT model studies \cite{vac, lars, schafwag12, chatmoh1,  guptiw, vkkr12, vkkt13, Rai},~where the QM model parameters are fixed using the curvature masses of the mesons in the presence of fermionic  vacuum fluctuations,~have shown that the quark one-loop vacuum fluctuations cause very large smoothing effect on the temperature and chemical potential variations of the non-strange, strange condensates, their derivatives and the mass $m_{\sigma}$ of the scalar $\sigma$ meson.~With the parameters obtained using the physical point $\pi,K,\eta \text{ and } \eta'$ masses,~the 2+1 flavor QMVT and RQM model order parameters,~their derivatives and phase diagrams in the $\mu-T$ plane,~have been compared extensively in Ref.\cite{vkkr22,skrvkt24} and it has been shown that the critical end point (where the crossover transition turns second order),~shifts to very large chemical potentials and smaller temperatures in the curvature mass parametrized QMVT model where the chiral crossover transition  on the temperature axis at $\mu=0$,~occurs at a noticeably larger  pseudo-critical temperature $T_{c}^{\chi}$ in comparison to the RQM model.~Recall that the smoothing effect of the quark one-loop vacuum fluctuation,~on the chiral transition  is moderate in the RQM model \cite{skrvkt24} because on-shell method of renormalizing the parameters in the RQM model,~generates  a significantly stronger axial $U_A(1)$ anomaly 't Hooft coupling $c$ while  the explicit chiral symmetry breaking 
strength $h_{x}$ ($h_{y}$) in the non-strange (strange) direction becomes weaker  by a small amount (a relatively large amount).~Note that the $c,h_{x} \text{ and } h_{y}$ do not change in the QMVT model at physical point from their tree level values obtained in the QM model.

\begin{table*}[!htbp]
\caption{Defining the reduction of $\pi,\ K$ meson starred masses as  $\frac{m_{\pi}^*}{m_{\pi}}= \frac{m_{K}^*}{m_{K}}=\beta \le 1$ such that $\frac{m_{\pi}^*}{m_{K}^*}=\frac{m_{\pi}}{m_{K}}$ with the $m_{\pi}=138$ and $m_{K}=496$ MeV at the physical point,~one gets direct path to reach the chiral limit \cite{Schaefer:09}.~The $f_{\pi}, \ f_{K}$ for a given $\beta$ fraction is obtained by using the $m_{\pi}^*$, $m_{K}^*$ for that $\beta$ in the Eq.~(\ref{fpi}) and Eq.~(\ref{fk}) while the Eq.~(\ref{Meta}) gives the corresponding $M_{\eta}$.~The experimental values $(m_{\eta},m_{\eta^{\prime}}) = (547.5,957.8)$ MeV give the $M_{\eta}$-Expt=1103.22 MeV while the infrared regularized U(3)  ChPT expression in the  Eq.~(\ref{Meta}) gives $M_{\eta}$=1035.55 MeV for the physical $m_{\pi}, m_{K}$.~The RQM$\lbrace \text{QMVT} \rbrace$ model parameters presented here are $\lambda_{20},c_{0},h_{x0}^{*},h_{y0}^{*}$ $\lbrace \lambda_{2},c,h_{x}^{*},h_{y}^{*} \rbrace$.}
\label{tab:table1}
\begin{tabular}{p{0.08\textwidth}| p{0.08\textwidth}|  p{0.08\textwidth}| p{0.15\textwidth}| p{0.13\textwidth} p{0.15\textwidth}| p{0.16\textwidth} p{0.18\textwidth}} 
\hline
$ \beta  $ & $f_{\pi} (\text{MeV})$ & $f_{K} (\text{MeV})$ & $\text{M}_{\eta}$ : (MeV) & $\lambda_{20} \ \lbrace  \lambda_{2} \rbrace$ & $ c_{0} \ \lbrace  c \rbrace (\text{MeV})$& $ h^*_{x0} \lbrace h^*_{x} \rbrace (\text{MeV}^3)$ & $ h^*_{y0} \lbrace h^*_{y} \rbrace (\text{MeV}^3)$  \\
\hline
1 &92.9737&113.2635 &1035.55 & 35.47  $\lbrace 54.46 \rbrace$ &7390.50  $\lbrace 3913.39 \rbrace$ & $(119.79)^3 \lbrace (120.98)^3 \rbrace$ & $(324.39)^3 \lbrace (336.65)^3 \rbrace $ \\
1 &92.9737&113.2635 & 650.43 & 90.81  $\lbrace 83.76 \rbrace$ &0  $\lbrace 0 \rbrace$ & \ \ \ \ \ \ \ \ \ "  & \ \ \ \ \ \ \ \ \ "   \\
0.5&92.7862&102.1041&865.99 &-12.95 $\lbrace 16.59 \rbrace$  &7705.88 $\lbrace 4241.24  \rbrace$ &$(75.98)^3 \lbrace (76.16)^3 \rbrace $  &$(197.66)^3 \lbrace (204.63)^3 \rbrace $ \\
0.4&91.7930&98.6312&849.04 &-22.45 $\lbrace 8.18 \rbrace$  &7808.30 $\lbrace 4448.38  \rbrace$ &$(65.30)^3 \lbrace (65.40)^3 \rbrace $  &$(169.62)^3 \lbrace (174.27)^3 \rbrace $ \\
0.3642 &91.3992&97.3737 &844.29 &-26.01 $\lbrace 4.95 \rbrace$  &7843.53 $\lbrace  4528.40 \rbrace$ &$(61.27)^3 \lbrace (61.35)^3 \rbrace $ &$(159.12)^3 \lbrace (163.01)^3 \rbrace $ \\
0.3 &90.6629& 95.1438 &837.31&-32.64$\lbrace -1.17 \rbrace$& 7901.53 $\lbrace 4675.25  \rbrace$ &$(53.71)^3
\lbrace (53.76)^3 \rbrace $&$(139.46)^3 \lbrace (142.1)^3 \rbrace $ \\
0.2028 &89.5459& 91.9877 &830.14&-43.33$\lbrace-11.31 \rbrace$& 7969.22 $\lbrace 4890.09  \rbrace$ &$(41.22)^3
\lbrace (41.24)^3 \rbrace $&$(107.02)^3 \lbrace (108.19)^3 \rbrace $\\ 
0.15&88.9846&90.4869&827.71&-49.53$\lbrace-17.32\rbrace$ &7992.99 $\lbrace4992.98 \rbrace$&$(33.65)^3
\lbrace (33.66)^3 \rbrace $&$(87.39)^3 \lbrace (87.996)^3 \rbrace $\\  
0.007246&88.0055&88.0129&825.049&-72.80$\lbrace -40.35 \rbrace$&8012.25$\lbrace5156.06\rbrace$ &$(4.448)^3
\lbrace (4.448)^3 \rbrace $&$(11.562)^3 \lbrace (11.561)^3 \rbrace $\\
\hline
1 &92.9737&113.2635 &$\text{M}_{\eta}$:Expt=1103.22 & 31.19  $\lbrace 47.93 \rbrace$ &7962.25  $\lbrace 4785.76 \rbrace$ & $(119.79)^3 \lbrace (120.98)^3 \rbrace$ & $(324.39)^3 \lbrace (336.65)^3 \rbrace $ \\
\hline
\end{tabular}
\end{table*}
\begin{table*}[!htbp]
\caption{The scalar $\sigma$ meson mass dependent parameters $\lambda_{10} \ \text{and} \ m_{0}^2$ $\lbrace \lambda_{1} \ \text{and} \ m^2 \rbrace$ of the RQM $\lbrace \text{QMVT} \rbrace$ model,~are presented in this table for the $m_{\sigma}=400$ MeV and $m_{\sigma}=530$ MeV.}
\label{tab:table2}
\begin{tabular} {p{0.07\textwidth}| p{0.07\textwidth} | p{0.08\textwidth}| p{0.10\textwidth} | p{0.15\textwidth} p{0.19\textwidth}| p{0.15\textwidth} p{0.19\textwidth}}
\hline	 
&  &  &  &   $m_{\sigma}=400 \ (\text{ MeV })$ & &  $m_{\sigma}=530 \ (\text{ MeV })$ &   \\
\hline
$ \beta  $ & $f_{\pi} (\text{MeV})$ & $f_{K} (\text{MeV})$ & $\text{M}_{\eta}$ : (MeV) &  $\lambda_{10} \ \lbrace  \lambda_{1} \rbrace$ & $ m_{0}^{2} \ \lbrace  m^2 \rbrace (\text{MeV}^2)$&  $\lambda_{10} \ \lbrace  \lambda_{1} \rbrace$ & $ m_{0}^{2} \ \lbrace  m^2 \rbrace  (\text{MeV}^2)$  \\
\hline
1 &92.9737&113.2635 & 1035.55 &  1.27 $\lbrace -12.18 \rbrace$ &$(447.71)^2 \lbrace (229.88)^2\rbrace $ & 4.499 $\lbrace -8.29 \rbrace$ &$(378.99)^2 \lbrace-(123.97)^2\rbrace $  \\
1 &92.9737&113.2635 & 650.43 &  -27.68 $\lbrace -27.65 \rbrace$ &$-(154.53)^2 \lbrace -(251.91)^2\rbrace $ &   -24.18 $\lbrace -23.65 \rbrace$ & $-( 292.08)^2 \lbrace -(365.47)^2\rbrace $ \\
0.5&92.7862&102.1041&865.99 & 20.19 $\lbrace 5.15 \rbrace$  & $(217.28)^2\lbrace -(284.53)^2 \rbrace$ &  23.96 $\lbrace 9.32 \rbrace$  &$-( 93.16)^2 \lbrace -(377.77)^2\rbrace $ \\ 
0.4&91.7930&98.6312&849.04 & 24.26 $\lbrace 9.34 \rbrace$  & $(169.93)^2\lbrace -(301.78)^2 \rbrace$ &  28.37 $\lbrace 13.70 \rbrace$  &$-( 169.47)^2 \lbrace -(390.01)^2\rbrace $ \\ 
0.3642 &91.3992&97.3737 &844.29 &25.78 $\lbrace 10.93 \rbrace$  &$(152.19)^2 \lbrace 
-( 305.97)^2 \rbrace$ & 30.04 $\lbrace 15.37 \rbrace$ &$-(187.68)^2 \lbrace 
-(393.02)^2 \rbrace$ \\
0.3 &90.6629& 95.1438 &837.31& 28.6$\lbrace 13.91 \rbrace$&  $ (118.4)^2\lbrace -(311.31)^2 \rbrace$ & 33.15$\lbrace 18.51 \rbrace$&$-( 214.51)^2 \lbrace -(396.88)^2 \rbrace $\\
0.2028 &89.5459& 91.9877 &830.14& 33.04 $\lbrace 18.67 \rbrace$&  $ (55.17)^2\lbrace -(314.96)^2 \rbrace$ & 38.08 $\lbrace 23.51 \rbrace$&$-( 244.73)^2 \lbrace -(399.56)^2 \rbrace $ \\
0.15&88.9846&90.4869&827.71& 35.52 $\lbrace 21.34 \rbrace$&  $-(36.92)^2 \lbrace -(315.17)^2 \rbrace$&  40.75 $\lbrace 26.32 \rbrace$&  $-(254.90)^2 \lbrace -(399.79)^2 \rbrace$ \\
0.007246&88.0055&88.0129&825.049&43.92 $\lbrace 30.1 \rbrace$&$-(84.02)^2\lbrace -(312.86)^2\rbrace$&49.87 $\lbrace  35.30 \rbrace$&$-(275.99)^2\lbrace -(397.91)^2\rbrace$\\
\hline
1 &92.9737&113.2635 &$\text{M}_{\eta}$:Expt  &  1.09 $\lbrace -8.71 \rbrace$ &$(444.12)^2 \lbrace (280.22)^2\rbrace $ & 4.32 $\lbrace -4.85 \rbrace$ &$(375.01)^2 \lbrace (103.06)^2\rbrace $  \\
\hline
\end{tabular}
\end{table*}

It is worthwhile to perform the chiral limit study and draw the Columbia plot in the curvature mass parametrized QMVT model where the infrared regularized U(3) ChPT is employed to determine the ($m_{\pi},m_{K}$) dependent numerical values of the $f_{\pi},f_{K} \text{ and } M_{\eta}^2$ when one reduces the $m_{\pi}$ and $m_{K}$ while moving away from the physical point.~The softening effect of the fermionic vacuum fluctuations in the QMVT model needs to be numerically quantified by calculating its effect on the location of the tricritical point (TCP) at $\mu=0$ in the light chiral limit and the value of the pion (light quark) critical mass $m_{\pi}^{c}$ ($m_{ud}^{c}$) beyond which the chiral transition is a smooth crossover on the three flavor chiral limit line ($m_{\pi}=m_{K}$  i.e.  $m_{ud}=m_{s}$ ) in the Columbia plot.~In order to see how the chiral limit results are qualitatively and quantitatively influenced by the different methods of treating the quark one-loop vacuum fluctuation when it gets incorporated in the QM Model,~the $\mu-m_{K}$, $m_{\pi}-m_{K}$ and the light-strange quark mass $\mu-m_{s}$, $m_{ud}-m_{s}$ planes of the Columbia plots,~the $m_{K}^{TCP}(m_{s}^{TCP})$ location of the TCP on the $m_{K}(m_{s})$ axis where the $m_{\pi}(m_{s})=0$  and the $m_{\pi}^{c}$ ($m_{ud}^{c}$) obtained for the QMVT model calculation in the present work will be compared with the corresponding results of the e-MFA:QM model FRG study in the Ref.~\cite{Resch} and the very recent RQM model chiral limit study reported  in the Ref.~\cite{vkt25}.
\begin{figure*}[htb]
\subfigure[]{
\label{fig1a} 
\begin{minipage}[b]{0.49\textwidth}
\centering
\includegraphics[width=\linewidth]{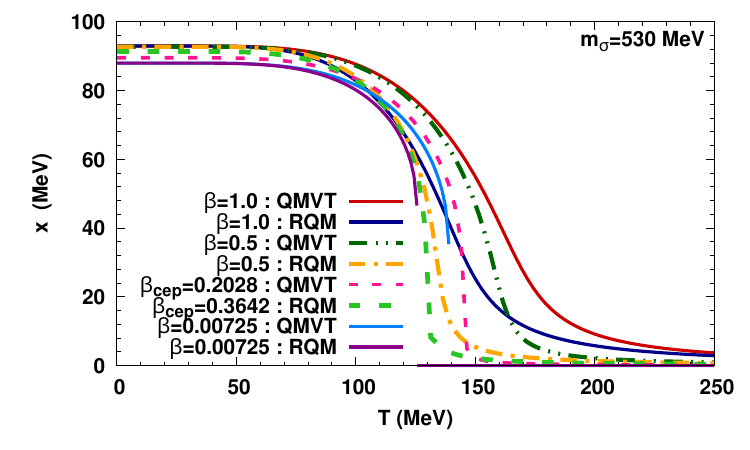}
\end{minipage}}
\hfill
\subfigure[]{
\label{fig1b} 
\begin{minipage}[b]{0.49\textwidth}
\centering 
\includegraphics[width=\linewidth]{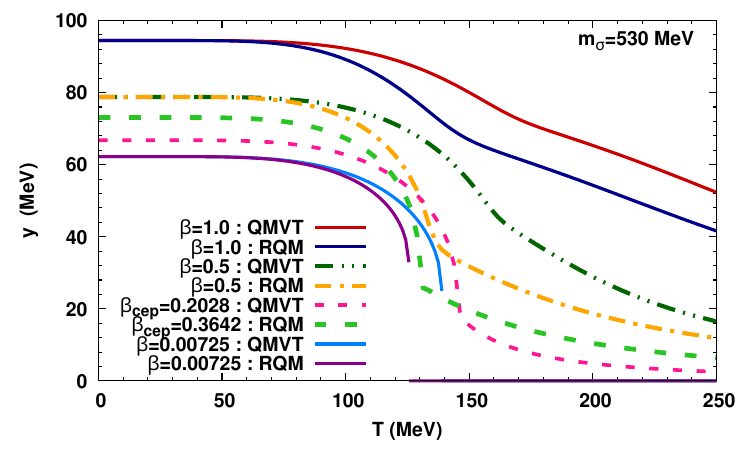}
\end{minipage}}
\caption{Temperature variations of $\text{x}$ and $\text{y}$ [(a) and (b)] at $\mu=0$ for the  ratio $\frac{m_{\pi}^*}{m_{\pi}}= \frac{m_{K}^*}{m_{K}}=\beta=1,0.5,0.00725.$The critical end point ratio $\beta_{\text{cep}}=0.2028(0.3642)$ in the QMVT (RQM) model. (a) Non-strange chiral condensate (b) Strange chiral condensate.}
\label{fig:mini:fig1} 
\end{figure*}
\begin{figure*}[htb!]
\subfigure[]{
\label{fig2a} 
\begin{minipage}[b]{0.49\textwidth}
\centering
\includegraphics[width=\linewidth]{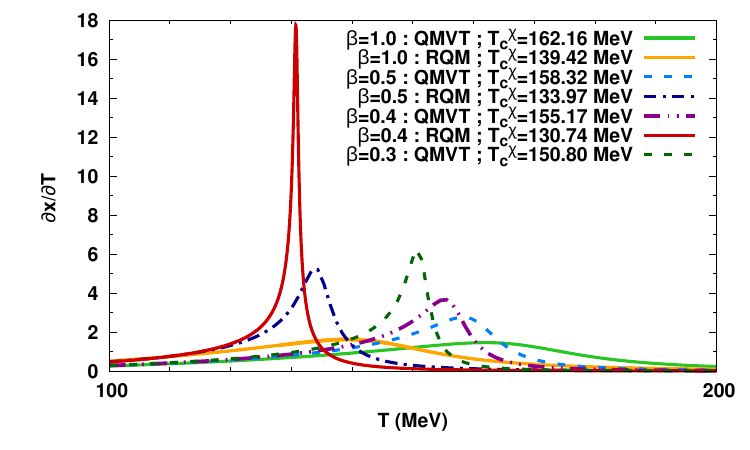}
\end{minipage}}
\hfill
\subfigure[]{
\label{fig2b} 
\begin{minipage}[b]{0.49\textwidth}
\centering 
\includegraphics[width=\linewidth]{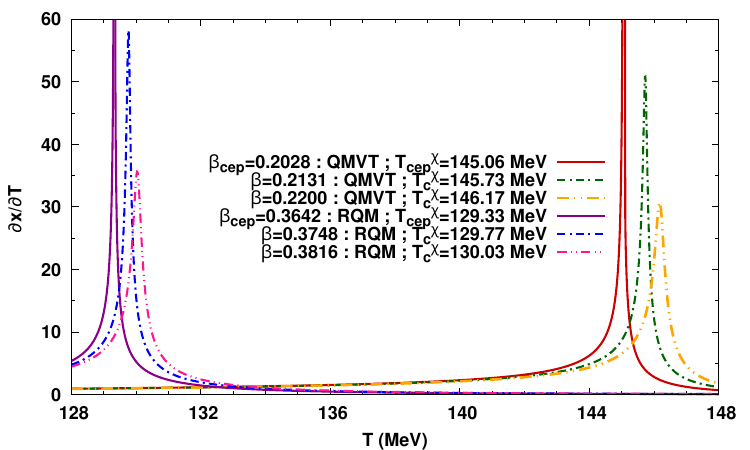}
\end{minipage}}
\caption{(a)Temperature variations of the light condensate  derivative $\partial \text{x}/ \partial \text{T}$ in the QMVT model  at $\mu=0$ for $m_{\sigma}=530$ MeV and $\frac{m_{\pi}^*}{m_{\pi}}= \frac{m_{K}^*}{m_{K}}=\beta=1,0.5,0.4, 0.3$~are compared with the corresponding values obtained in the RQM model. (b) The $\partial \text{x}/ \partial \text{T}$ temperature plots for the QMVT model cep ratio $\beta_{cep}=0.2028 \ \{(m_{\pi}^*\ ,m_{K}^*)=(27.99,100.59)  \text{ MeV }\}$ and in its close proximity when $\beta=0.2131 \ \{(m_{\pi}^*\ ,m_{K}^*)=(29.41,105.70) \text{ MeV} \}    \text{ and }  \beta=0.2200 \ \{(m_{\pi}^*\ ,m_{K}^*)=(29.92,109.12) \text{ MeV}\} $,~show the respective peaks at the transition temperatures of $T_{cep}^{\chi}=145.06$, $T_{c}^{\chi}=145.73 \text{ and } 146.17$ MeV.~The divergent peak in the RQM model $\partial \text{x}/ \partial \text{T}$ plot,~occurs at a significantly smaller $T_{cep}^{\chi}=129.33$ MeV for a noticeably larger ratio $\beta_{cep}=0.3642 \  \{(m_{\pi}^*\ ,m_{K}^*)=(50.26,180.64) \ \text{MeV}\}$.~Adjacent RQM model plots for $\beta=0.3748 \ \{(m_{\pi}^*\ ,m_{K}^*)=(51.72,185.90)\text{ MeV}\} \ \ \text{and} \ \beta=0.3816 \ \{(m_{\pi}^*\ ,m_{K}^*)=(52.66,189.27) \text{ MeV} \}$ show,~respective crossover transition peaks at $T_{c}^{\chi}=129.77 \text{ and } 130.03 \text{ MeV}$.}
\label{fig:mini:fig2} 
\end{figure*}
\subsection{\bf{QMVT versus RQM model order parameters and its derivatives in the path to chiral limit}}
\label{subsec:Chlmt}

The light chiral limit in the QMVT (RQM) model is given by $h_{x}=0$ ($h_{x0}=0$)  while the strange  chiral limit is given by $h_{y}=0$ ($h_{y0}=0$).~The reduced chiral  symmetry breaking strengths  $h_{x}^*$ and $h_{y}^*$ ($h_{x0}^*$ and $h_{y0}^*$) and the corresponding QMVT (RQM) model parameters $\lambda_{2},c,\lambda_{1},m^2 \ (\lambda_{20},c_{0},\lambda_{10},m_{0}^2)$ for the chiral limit study,~are calculated by  using the  infrared regularized U(3) ChPT described in the section \ref{subsec:Chpt}.~The changes in  pion,~kaon decay constants $f_{\pi}$ and $ f_{K}$ are calculated by putting the values of the reduced pion,~kaon masses $m_{\pi}^*$ and $m_{K}^*$ respectively into the Eq.~(\ref{fpi}) and Eq.~(\ref{fk}) while $\text{M}_{\eta}$ is calculated using the  Eq.(\ref{Meta}).~The ratio $\frac{m_{\pi}^*}{m_{\pi}}= \frac{m_{K}^*}{m_{K}}=\beta$ is kept fixed while the starred $\pi$ and $K$ masses are reduced.~The above choice gives us  a direct path from the physical point to the chiral limit by just tuning  only the fraction  $\beta$ instead of randomly changing both the $m_\pi$ and $m_K$ and ensures that the $\frac{m_{\pi}^*}{m_{K}^*}$ remains equal to the ratio $\frac{m_{\pi}}{m_{K}}$ \cite{Schaefer:09} at the physical point where the ChPT parameter $q=\frac{2 \ m_s}{m_u+m_d}=\frac{ m_s}{m_{ud}}$ as the ratio of the strange $m_{s}$ to the average of light quark mass $m_{ud}$,~also gets fixed.~The above defined fraction $\beta$,~also facilitates the comparison of our results with those of the Refs.~\cite{Resch,Schaefer:09,vkt25} for the  analogous fractions.~The reduced
light and strange chiral symmetry breaking strengths in the QM/QMVT model are defines as $h_{x}^{*}=(m_{\pi}^{*})^{2} \ f_{\pi}$ and $h_{y}^{*}=\lbrace\sqrt{2} \ f_K \ (m_{K}^{*})^{2}-\frac{f_{\pi}}{\sqrt{2}} \ (m_{\pi}^{*})^{2}\rbrace$.~The analogous definitions in the RQM model are : $h_{x0}^{*}=(m_{\pi,c}^{*})^{2} \ f_{\pi}$ and $h_{y0}^{*}=\lbrace\sqrt{2} \ f_K \ (m_{K,c}^{*})^{2}-\frac{f_{\pi}}{\sqrt{2}} \ (m_{\pi,c}^{*})^{2}\rbrace$ where the 
reduced pion,~kaon curvature masses $m_{\pi,c}^{*} \text{ and } m_{K,c}^{*} $ are found by putting the reduced pion,~kaon mass $m_{\pi}^{*} \text{ and } m_{K}^{*} $ values respectively in the expressions Eq.~(\ref{mpicr}) and Eq.~(\ref{kcr}).

The RQM$\lbrace{\text{QMVT}\rbrace}$ model parameters $\lambda_{20},c_{0},h_{x0},h_{y0}$ $\lbrace \lambda_{2},c,h_{x},h_{y} \rbrace$ for the $f_{\pi},f_{K},M_{\eta}$ when the $\beta=1,0.5,0.4,0.3642,0.3,0.2028,0.15,0.007246 $ are presented in the Table \ref{tab:table1} and the Table \ref{tab:table2} contains the corresponding RQM $\lbrace{\text{QMVT}\rbrace}$  model parameters $\lambda_{10},m_{0}^2$ $\lbrace \lambda_{1},m^2 \rbrace$ for the $\sigma $ masses $m_{\sigma}=400$ MeV and $m_{\sigma}=530$ MeV.~The $m_{\pi}$=138.0 MeV,~$m_{K}$=496.0 MeV for $\beta=1$ at the physical point.~The RQM $\lbrace{\text{QMVT}\rbrace}$ model parameters in the absence of the axial $U_A(1)$ anomaly i.e. $c=0$  at the physical point,~are presented in the second row of the Tables \ref{tab:table1} and \ref{tab:table2} whose  last row presents the parameters for the $M_{\eta}:\text{Expt}$ in the presence of the axial $U_A(1)$ anomaly and $f_{\pi},f_{K}$ are the same as in the first row as calculated from the ChPT input.~For the physical point experimental masses $(m_{\eta},m_{\eta^{\prime}}) = (547.5,957.8)$ MeV,~the input of $M_{\eta}=\sqrt{(m_{\eta}^2+m_{\eta^{\prime}}^2)}=1103.22$ MeV,~is termed as $M_{\eta}:\text{Expt}$ for which the calculation of parameters in the
curvature mass parametrized QMVT model,~give $(m_{\eta},m_{\eta^{\prime}}) = (538.54,962.85)$ MeV in the output while the calculation of parameters in the on-shell renormalized  RQM model,~give $(m_{\eta},m_{\eta^{\prime}}) = (527.82,968.76)$ MeV in the output.~Since the obtained $M_{\eta}$ in the output  of the QMVT (RQM) model is the same as in the input,~the above results are consistent.~One gets a slightly smaller value of the $M_{\eta}=1035.55$ MeV when the ChPT expression  in Eq.(\ref{Meta}) is used with the physical point $m_{\pi}$,~$m_{K}$.~For $M_{\eta}=1035.55$ MeV in the input,~the RQM model parameters give $(m_{\eta},m_{\eta^{\prime}}) = (520.62,895.17)$ MeV in the output while the QMVT model parameters give $(m_{\eta},m_{\eta^{\prime}}) = (528.76,890.79)$ MeV in the output and the  output,~input values for $M_{\eta}$ are same in both the models.

The QMVT and RQM model temperature variations of the  light and strange condensate $x$ and $y$ have been been compared respectively in the  Fig.~\ref{fig1a} and Fig.~\ref{fig1b} when $\beta=1.0,0.5,0.3642,0.2028 \text{ and } 0.00725$ for the $m_{\sigma}=530$ MeV.~When ChPT expression of $M_{\eta}$ is used as input to determine parameters of the QMVT and RQM models,~the $x$ ($y$) plot at the physical point $\beta=1$,~overlaps (not shown here),~as  reported in a very recent RQM model work in the Ref.~\cite{vkt25},~with the corresponding temperature plot obtained after taking  the $M_{\eta}:\text{Expt}$ in input for finding the model parameters whereas the $f_{\pi},f_{K}$ are the same as in the first row of Tables \ref{tab:table1} and \ref{tab:table2}.~The light and strange condensate temperature variations in the Fig.~\ref{fig1a} and Fig.~\ref{fig1b},~show a very smooth and delayed melting in the QMVT model for the $\beta=1.0 \text{ and } 0.5$ while the corresponding melting in the RQM model is sharper and occurs early on the temperature axis.~This is evident from the temperature variation of the derivative $\partial x / \partial T$ in the  Fig.~\ref{fig2a} where the chiral crossover transition in the QMVT model occurs at the
pseudo-critical temperature (defined as the peak of the $\partial x / \partial T$ temperature variation) of $T_{c}^{\chi}=162.16$ MeV for $\beta=1$ and $T_{c}^{\chi}=158.32$ MeV for $\beta=0.5$ while a relatively sharper chiral crossover transition,~with a higher peak in the $\partial x / \partial T$ temperature plot,~occurs in the RQM model at about 23-24 MeV smaller transition temperature of $T_{c}^{\chi}=139.42$ MeV for $\beta=1$ and $T_{c}^{\chi}=133.97$ MeV for $\beta=0.5$ .~The $\partial x / \partial T$ peak height in the QMVT model respectively for the $\beta=0.4$ and $\beta=0.3$ in the Fig.~\ref{fig2a} turn out to be about five and three times smaller than the height of the $\partial x / \partial T$ peak in the RQM model for the case of $\beta=0.4$ .~The reason of the above pattern lies in the fact that the $\beta=0.4$ is close to the RQM model $\beta_{cep}|_{(m_{\sigma}=530\text{ MeV}:\text{ RQM})}=0.3642 \text{ for } (m_{\pi,cep}^*,m_{K,cep}^*)=(50.26,~180.64) $ MeV where the peak in the $\partial x / \partial T$ temperature plot of the Fig.~\ref{fig2b},~diverges at the $T_{c}^{\chi}=129.33$ MeV and the crossover transition ends at the second order $Z_{2}$ critical point to become first order transition for the $\beta<\beta_{cep}$.~The RQM model  $\partial x / \partial T$ plots for  the $\beta=0.3748$ and $\beta=0.3816$ close to the $\beta_{cep}|_{(m_{\sigma}=530\text{ MeV}:\text{ RQM})}=0.3642$ are also shown in the Fig.~\ref{fig2b} to have a perspective of the divergence.~The fraction $\beta$ needs significant reduction to get the  $\beta_{cep}|_{(m_{\sigma}=530\text{ MeV}:\text{QMVT})}=0.2028 \text{ for } (m_{\pi,cep}^*,m_{K,cep}^*)=(27.99,~100.59) $ MeV in the QMVT model when the peak height of the $\partial x / \partial T$ plot diverges at larger $T_{c}^{\chi}=145.06$ MeV in the  Fig.~\ref{fig2b} where the adjacent QMVT model $\partial x / \partial T$ plots are shown for the $\beta=0.2131$ and $\beta=0.2200$ also.~The Fig.~\ref{fig1a} and \ref{fig1b} show the light and strange order parameter temperature variations also for the $\beta_{cep}$ and $\beta=0.00725$ where the order parameters have a very large first order gap when one reaches very close to the chiral limit.~Note that for the smaller $\beta<\beta_{cep}$,~the  difference in the QMVT and RQM model critical temperatures for the chiral transition decreases from about 24 MeV for the $\beta=1.0 \text{ and }0.5$ to 16.35 MeV for $\beta=0.15$ and 13.35 MeV for $\beta=0.00725$.~In the exact chiral limit $f_{\pi}=f_{K}$,~the  $m_{\eta}=0$ and the heavy $m_{\eta^{\prime}} \equiv 825$ MeV signifies the $U_A(1)$ anomaly for both the QMVT model and RQM model.~The problem of the loss of the spontaneous chiral symmetry breaking ( SCSB) when the $m_{\pi},m_{K} \rightarrow  0$  in the chiral limit for the moderate $m_{\sigma}=400-800$ MeV,~gets cured after using the infrared regularized U(3) ChPT scaling \cite{Herpay:05} of the  $m_{\pi},m_{K},M_{\eta}^2$.

\begin{figure*}[htb]
	\subfigure[]{
		\label{fig3a} 
		\begin{minipage}[b]{0.49\textwidth}
			\centering
			\includegraphics[width=\linewidth]{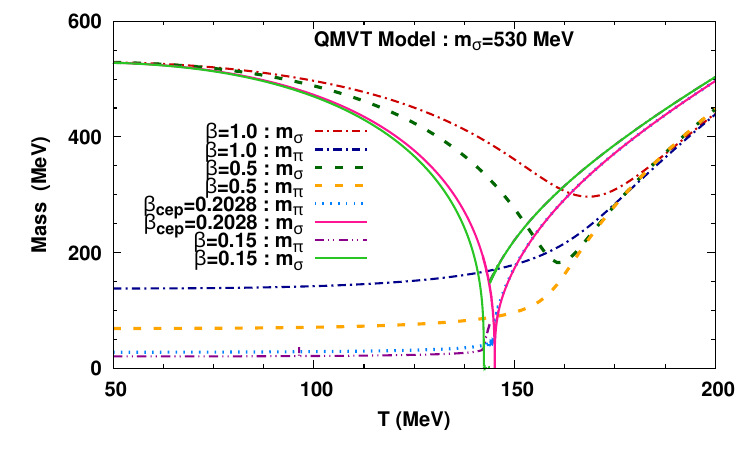}
	\end{minipage}}
	\hfill
	\subfigure[]{
		\label{fig3b} 
		\begin{minipage}[b]{0.49\textwidth}
			\centering 
			\includegraphics[width=\linewidth]{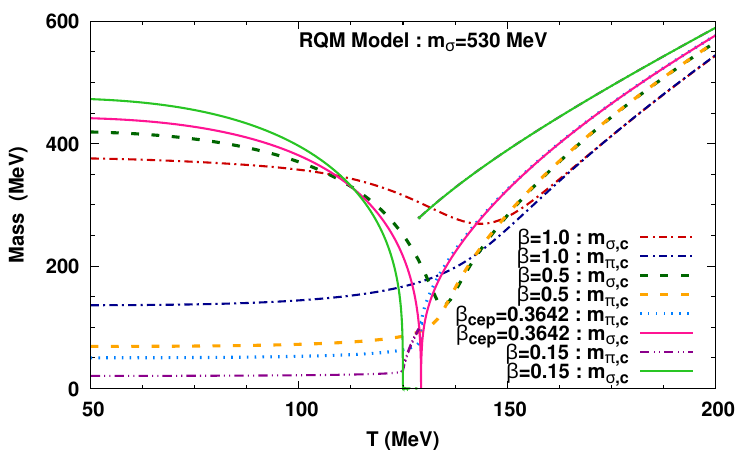}
	\end{minipage}}
	\caption{Temperature variations of the $\pi, \ \sigma$ meson masses $m_{\pi} \text{ and } m_{\sigma}$  at $\mu=0$ for $m_{\sigma}=530$ MeV and $\frac{m_{\pi}^*}{m_{\pi}}= \frac{m_{K}^*}{m_{K}}=\beta=1,0.5,0.3642,0.2028,0.015$ are shown in the left panel (a) for the QMVT model and right panel (b) for the RQM model.~The curvature and pole masses are same in the QMVT model while the curvature mass $m_{\pi,c} (m_{\sigma,c})$  of the $\pi (\sigma)$ meson is smaller (significantly smaller) than its pole mass of $m_{\pi} (m_{\sigma})=138 (530) \text{ MeV}$  in the RQM model.~When the $m_{\sigma}$ becomes exactly zero in its temperature variation,~it gives the signature identification of the cep respectively at the $\beta_{cep}=0.2028$ for the QMVT model in (a) and the $\beta_{cep}=0.3642$ for the RQM model in (b).~Strong first order mass gap is seen in the $m_{\sigma}$  plot for the  $\beta=0.015<\beta_{cep}$.}
	\label{fig:mini:fig3} 
\end{figure*}

~It is worthwhile to compare the $\beta_{cep}$ for the case of the $m_{\sigma}=400$ MeV also, though the order parameter and its derivative  plots are not shown.~The critical end point fraction in the QMVT model is found at the  $\beta_{cep}|_{(m_{\sigma}=400\text{ MeV}:\text{QMVT})}=0.2215 \text{ for } (m_{\pi,cep}^*,m_{K,cep}^*)=(30.57,~109.88) $ MeV  when the $m_{\sigma}=400$ MeV while the similar cep in the RQM model is found at the  $\beta_{cep}|_{(m_{\sigma}=400\text{ MeV}:\text{RQM})}=0.3795 \text{ for } (m_{\pi,cep}^*,m_{K,cep}^*)=(52.38,~188.25) $ MeV.~It is  known that the strength of the chiral transition becomes stronger for smaller scalar $\sigma$ meson masses.~The above feature gets verified as we are finding the $\beta_{cep}|_{(m_{\sigma}=400\text{ MeV}:\text{QMVT})}=0.2215 > \beta_{cep}|_{(m_{\sigma}=530\text{ MeV}:\text{QMVT})}=0.2028$ in the QMVT model while we find  the $\beta_{cep}|_{(m_{\sigma}=400\text{ MeV}:\text{RQM})}=0.3795 > \beta_{cep}|_{(m_{\sigma}=530\text{ MeV}:\text{RQM})}=0.3642$ for the RQM model.~The $\beta_{cep}$ computed using the large $N_{c}$ standard U(3) ChPT input set $\text{M}_{\eta}$-II in the RQM model gives slightly different $\beta_{cep}|_{(m_{\sigma}=530\text{ MeV}:\text{RQM};\text{M}_{\eta}-\text{II})}=0.36887$.

~It is important to compare  the QMVT model result of the present work with the similar result of the Ref.~\cite{Resch} where the effect of quark one-loop vacuum fluctuation was computed in the e-MFA:QM Model FRG work under the LPA by switching off  the quantum and thermal fluctuations of the mesons in the functional renormalization group flow equation.~Recall that in the Ref.~\cite{Resch},~they reduce the  non-strange and strange direction  chiral symmetry breaking source strengths defined as $j_{x} \text{ and } j_{y}$ in their work by tuning a single parameter  $\alpha=(0,0.04,0.17,1)$ but corresponding to each of the above reduction in the $\alpha$,~they respectively adjust the initial action by changing the scale $\Lambda=(1143,1100,1000,700) \text{ MeV }$ such that the $f_{\pi}$ always remains fixed and the spontaneous chiral symmetry breaking does not get lost.~The crossover transition for the $\alpha=1$ at the physical point,~becomes second order for the critical  $\alpha_{c}=.04$ and one gets the first order transition for the $\alpha_{c}<.04$ when the $m_\sigma=530$ MeV in their work.~The parameter $\alpha$ of their study gives the related parameter $\beta=\sqrt{\alpha}$ in our work.~Comparing our QMVT model results with those of the Ref.~\cite{Resch},~we see that the second order $Z_{2}$ critical point in their e-MFA:QM model study occurs at  $\beta_{cep}|_{(m_{\sigma}=530\text{ MeV}:\text{eMFA;QM-FRG})}=\sqrt{\alpha_{c}}=0.2$ for  $(m_{\pi,cep}^*,m_{K,cep}^*)=(27.6,~99.2)$ MeV which is quite close to our QMVT model result of the $\beta_{cep}|_{(m_{\sigma}=530\text{ MeV}:\text{QMVT})}=0.2028$.~Thus the smoothing effect of fermionic vacuum fluctuation on the chiral transition is significantly large (and similar) in the QMVT model as well as the e-MFA:QM model FRG study of Ref.~\cite{Resch} where the extent  of  first order transition region is small.~The above softening effect is quite moderate in the  RQM model study which has already been reported very recently in the Ref.~\cite{vkt25} where one finds a noticeably larger extent of the first order region and quite a large $\beta_{cep}|_{(m_{\sigma}=530\text{ MeV}:\text{RQM})}=0.3642$ also.~Here it is relevant to point out that when the Dirac's sea contribution gets neglected under the s-MFA in the QM model,~the Fig.~(9) of the Ref.~\cite{Schaefer:09},~shows  that the crossover transition ends at the second order transition for the significantly larger fraction $\beta_{cep}=.488$ giving noticeably larger $\pi \text{ and } K$ masses $(m_{\pi,cep}^*,m_{K,cep}^*)=(67.34,~242.05)$ MeV where they have taken $m_{\sigma}=800\text{ MeV}$ in their calculation so that the SCSB is not lost.~The Table~(\ref{tab:table3}) shows the comparison of different results for the chiral limit studies done in the QMVT model,~RQM model,~e-MFA:FRG QM model and the s-MFA QM model.~The table shows the LQCD results also for the ready reference. 


 \subsection{\bf{Comparing the $m_{\sigma}$ towards the chiral limit}}
\label{subsec:sigma}

The temperature variation of the $\sigma \text{ and } \pi$ masses is also used for the analyzing the chiral transition as it is known that the $m_\sigma$ and $m_\pi$ become degenerate after the chiral transition.~Furthermore when the scalar mass $m_\sigma$ becomes zero in its temperature variation,~it signals identification of the second order Z(2) critical end point for the used model parameter set where the chiral crossover transition ends to become a second order transition.~The scalar $\sigma$ meson is a critical mode of the chiral transition and it is related to the correlation length as $\xi=1/m_{\sigma}$.~It has been argued and discussed at length in the Ref.\cite{Resch} that when chiral symmetry breaking light and strange source strengths $j_{x} \text{ and } j_{y}$ are reduced in going away from the physical point to make chiral limit study in the fixed UV scheme (where other parameters take the same value as at the physical point),~the mass $m_{\sigma}$ of the critical sigma mode increases and  the correlation length $\xi$ decreases to such an extent that the spontaneous breaking of chiral symmetry gets lost.~They salvaged the situation in their chiral limit study by proposing the ChPT motivated fixed-$f_{\pi}$ scheme where they heuristically adjust the initial effective action  to the larger scales ($\Lambda^{\prime} > \Lambda $) while reducing the source strengths $j_{x} \text{ and } j_{y}$ for smaller ($m_{\pi}, \ m_{K}$) away from the physical point such that the $f_{\pi}$ always retains its physical value and hence the SCSB is not lost for the scalar $\sigma$ mass range $m_\sigma=400-600$ MeV.~The changing scale ($\Lambda^{\prime} > \Lambda $) accounts for the change in parameters (which are kept same as at the physical point) when the explicit chiral symmetry breaking source strengths decrease.~Since we are changing the $f_{\pi}.f_{K} \text{ and } M_{\eta}$ away from the physical point in our chiral limit study according to the ($m_{\pi}, \ m_{K}$) dependent ChPT scaling relations,~it is interesting and relevant to compare the temperature variation of the $m_{\sigma}$ in our QMVT and RQM model study with the corresponding result in the Ref.\cite{Resch}.

\begin{figure*}[htb]
\subfigure[]{
\label{fig4a} 
\begin{minipage}[b]{0.49\textwidth}
			\centering
			\includegraphics[width=\linewidth]{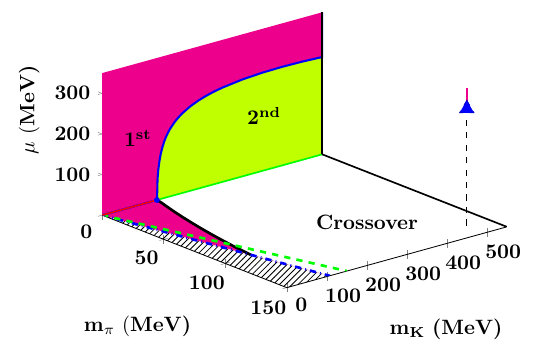}
	\end{minipage}}
	\hfill
	\subfigure[]{
		\label{fig4b} 
		\begin{minipage}[b]{0.49\textwidth}
			\centering 
			\includegraphics[width=\linewidth]{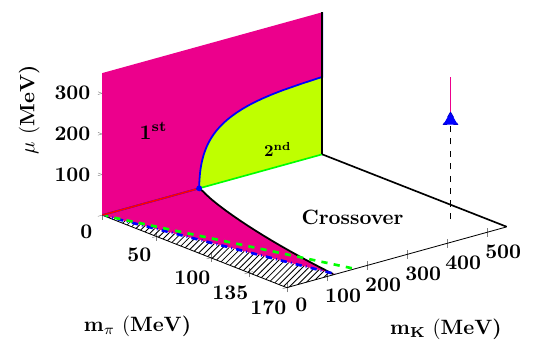}
	\end{minipage}}
\caption{~Columbia plot with the $U_{A}$(1) anomaly is presented in the left (right) panel (a) ((b)) for the QMVT (RQM) model when $m_{\sigma}=400$ MeV.~The $\mu-m_{K}$ ($m_{\pi}-m_{K}$) plane presents the chiral transition for the $m_{\pi}=0$ ($\mu=0$).~The second order $Z(2)$ critical solid black line demarcating the crossover from the first order region,~intersects the SU(3) symmetric $m_{\pi}=m_{K}$ green dash line respectively at the critical pion mass of $m_{\pi}^{c}=78.45$ and $m_{\pi}^{c}=134.16$ MeV in (a) for the QMVT model and (b) for the RQM model.~It terminates at the terminal pion mass of $m_{\pi}^{t}=98.15$ MeV in (a) for the QMVT model and $m_{\pi}^{t}=168.39$ MeV in (b) for the RQM model.~Solid blue line of the tricritical points that separates the second  and first order regions in the $\mu-m_{K}$ plane,~starts on the $m_K$ ($m_{\pi}=0$) axis from the blue dot respectively at the $m_{K}^{TCP}=137.2$ MeV in (a) for the QMVT model and $m_{K}^{TCP}=242.6$ MeV in (b) for the RQM model.~The blue dash line in the $m_{\pi}-m_{K}$ plane,~depicts the strange chiral limit $h_{y}=0$ and the $h_{y}$ is negative in the shaded unphysical region.~The black dash vertical line of crossover transition at the physical point,~ends at the critical end point with $\mu_{CEP}=288.03 \text{ MeV}$ in (a) for the QMVT model and  $\mu_{CEP}=243.29 \text{ MeV}$ in (b) for the RQM model.~The vertical red line depicts the first order transition.}
\label{fig:mini:fig4} 
\end{figure*}

~Note that the curvature and pole masses are same in the QMVT model whereas the curvature mass of the $\pi(\sigma)$ meson $m_{\pi,c} (m_{\sigma,c})=135.98 (376.74) \text{ MeV}$ 
is smaller (significantly smaller) than its pole mass of $m_{\pi} (m_{\sigma})=138 (530) \text{ MeV}$ for $\beta=1$ in the RQM model due to the on-shell renormalization of the parameters.~The QMVT model temperature variation of the $\sigma$ mass in the Fig.~\ref{fig3a},~begins from the same value $m_{\sigma}=530$ MeV for all the $\beta$ fractions whereas the RQM model curvature mass of the $\sigma$ meson in the  Fig.~\ref{fig3b} starts from the $m_{\sigma,c}=376.74,420.29,443.01 \text{ and } 474.75$ MeV respectively for the $\beta=1.0,0.5,0.3642 \text{ and }0.15$.~Thus we find that the vacuum curvature mass of the $\sigma$ meson in the RQM model,~is lowest at the physical point and it increases successively when one moves towards the chiral limit after reducing the ($m_{\pi},m_{K}$) by a fraction $\beta$ of their physical point value.~The behavior of the $m_{\sigma}$ in our chiral limit study for both the QMVT and RQM models,~stands in contrast to its behavior in the e-MFA-FRG, QM model study of the Ref.\cite{Resch} where in the aftermath of the successive increase in the scale $\Lambda=700,1000,1100 \text{ and } 1143$ corresponding to the successive decrease of the source strengths $j_{x} \text{ and } j_{y}$ governed by the $\alpha=\sqrt{\beta}=1.0,0.17,0.04 \text{ and } 0$,~the $\sigma$ mass registers a significant successive decrease towards the chiral limit as evident from the Fig.2(c) of the Ref.\cite{Resch}.~Here,~it is relevant to point out that the $\sigma$ mass in the e-MFA-FRG study is calculated from the curvature of the effective potential that gets evaluated by the FRG flow equation.~Recall that the curvature masses of mesons are used for fixing the parameters in the QMVT model and this context of incorporating quark one-loop vacuum correction is analogous to that of the e-MFA-FRG study in the Ref.\cite{Resch} where the vacuum value of the $m_{\sigma}$ decreases significantly towards the chiral limit while the QMVT model $m_{\sigma}$ does not change towards the chiral limit.~In complete contrast of the above two scenarios,~being lowest at the physical point and different from its pole mass $m_{\sigma}=530$ MeV,~the vacuum curvature mass $m_{\sigma,c}$ in the on-shell renormalized RQM model increases towards the chiral limit. 

The variation of the $m_{\sigma}$ with respect to the temperature is very smooth in the  Fig.~\ref{fig3a} of the QMVT model when $\beta=1.0 \text{ and } 0.5$ while the  $m_{\sigma,c}$   RQM model plot in the Fig.~\ref{fig3b} is sharper for the $\beta=0.5$,~though being quite smooth at the physical point $\beta=1.0$.~The respective temperature plots of the QMVT model $m_{\pi}$  and the RQM model $m_{\pi,c}$ in the Fig.~\ref{fig3a} and \ref{fig3b},~begin from their reduced value for $\beta<1$ and become degenerate respectively  with the $m_{\sigma}$ and $m_{\sigma,c}$  temperature variations for the corresponding $\beta$  after the chiral transition takes place.~The vacuum $\pi $ curvature mass from where the $m_{\pi,c}$ plots begin in the Fig.~\ref{fig3b} are,~  $m_{\pi,c}=135.98,68.75,50.17 \text{ and } 20.69$ in respective order for the $\beta=1.0,0.5,0.3642 \text{ and } 0.15$.~The $m_{\sigma}$ becomes zero in its temperature  variation in the Fig.~\ref{fig3a} when the $\beta_{cep}=0.2028$ signaling identification of the critical end point in the QMVT model.~The signature identification of the critical end point  results in the RQM model when the curvature mass of the $\sigma$ meson $m_{\sigma,c}$ becomes zero in its temperature variation in the Fig.~\ref{fig3b} for the $\beta_{cep}=0.3642$.~For very small $\beta=0.15$ close to the chiral limit,~the QMVT model $m_{\sigma}$ temperature plot in the Fig.~\ref{fig3a} and the RQM model  $m_{\sigma,c}$ temperature plot in Fig.~\ref{fig3b},~show a large mass gap that corresponds to the strong first order chiral transition.

\begin{figure*}[htb]
\subfigure[]{
\label{fig5a} 
\begin{minipage}[b]{0.49\textwidth}
			\centering
			\includegraphics[width=\linewidth]{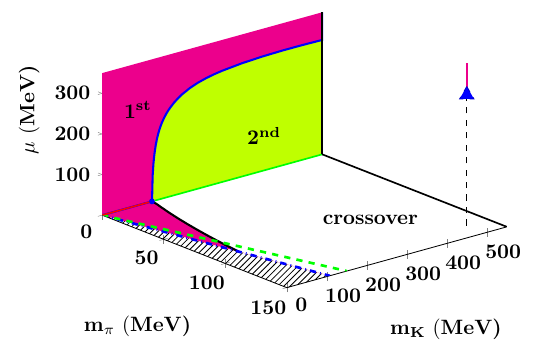}
	\end{minipage}}
	\hfill
	\subfigure[]{
		\label{fig5b} 
		\begin{minipage}[b]{0.49\textwidth}
			\centering 
			\includegraphics[width=\linewidth]{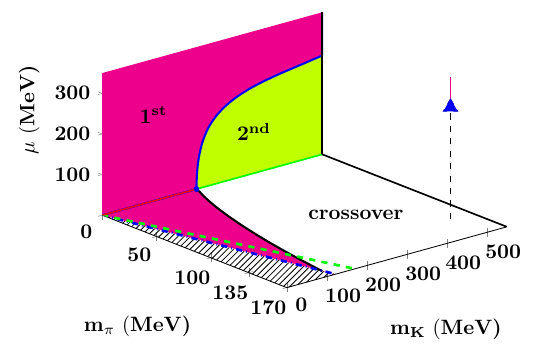}
	\end{minipage}}
\caption{~Left panel (a) (right panel (b)),~depicts the Columbia plot with the $U_{A}$(1) anomaly for the QMVT (RQM) model when $m_{\sigma}=530$ MeV.~The $m_{\pi}=0$ on $\mu-m_{K}$ plane while $\mu=0$ for $m_{\pi}-m_{K}$ plane.~The critical points,~lines,~first, second order,~crossover regions and other features are defined similar to the Fig.(\ref{fig:mini:fig4}).~The left panel (a) Columbia plot,~gives the $(m_{K}^{TCP},~m_{\pi}^{t},~m_{\pi}^{c}) \equiv (124.7,~90.03,~72.12)$ MeV for the QMVT model.~For the RQM Model,~Columbia plot in (b)  gives $(m_{K}^{TCP},~m_{\pi}^{t},~m_{\pi}^{c}) \equiv (235.7,~160.75,~128.38)$ MeV.~Vertical line of crossover transition at the physical point, terminates at the critical end point with $\mu_{CEP}=322.16 \text{ MeV}$ in (a) for the QMVT model and  $\mu_{CEP}=275.95 \text{ MeV}$ in (b) for the RQM model.}
\label{fig:mini:fig5} 
\end{figure*}
 
\subsection{\bf{Comparing various aspects of Columbia plots in the QMVT and RQM model for different $\bf{m_{\sigma}}$}}
\label{subsec:Colplot}
One needs different conditions other than simply reducing the  parameter $\beta$ to compute the distinct regions of the Columbia plot.~Implementing the light chiral limit
$m_{\pi}=0\implies h_{x}=h_{x0}=0$,~one gets the blue solid line in the vertical $\mu-m_{K}$ plane which is locus of the tricritical points that separates the first order from the second order transition region.~The $Z(2)$ critical end points (cep) in the horizontal  $m_\pi-m_{K}$ plane at  $\mu=0$,~are obtained after taking some numerical value for the $m_{\pi,cep}$ and then finding the corresponding $m_{K,cep}$ at which the crossover transition  for $m_{K}>m_{K,cep}$, turns second order to become first order for the $m_{K}<m_{K,cep}$.~The locus of the Z(2) critical end points,~gives the chiral critical line (denoted by the solid black line in the figures below) that separates the first order from the crossover transition region.~The condition $m_{\pi}=m_{K}$,~gives the $SU(3)$ symmetric line shown by the  green dash line in the figures below.~The strange chiral limit  $h_{y}=0=h_{y0}$ condition in the $m_\pi-m_{K}$ plane,~gives us the blue dash dotted line under which the strange direction chiral symmetry breaking strength $h_{y}$ becomes negative in the shaded area.

The QMVT model Columbia plots with the $U_{A}$(1) anomaly for the $m_{\sigma}=400$ MeV and $m_{\sigma}=530$ MeV are presented respectively in  the Fig.~(\ref{fig4a}) and Fig.~(\ref{fig5a}).~The Fig.~(\ref{fig4b}) and Fig.~(\ref{fig5b}) depict the corresponding RQM model Columbia plots for comparison.~The light chiral limit $m_{\pi}=0$ axis chiral transition for the  $m_{K} \ge 496$ MeV in all the above mentioned figures,~is of second order as in the \cite{Resch} and also predicted in the Ref. \cite{rob}.~The O(4) universality class second order transition line,~when the $m_{K}<496$ MeV in the QMVT model,~terminates at the blue dot of the tricritical point $m_{K}^{TCP}=137.2$ MeV for the $m_{\sigma}=400$ MeV and $m_{K}^{TCP}=124.7$ MeV for the $m_{\sigma}=530$ MeV respectively in the Fig.~(\ref{fig4a}) and Fig.~(\ref{fig5a}) where the transition turns first order and keep on becoming stronger till the chiral limit is reached.~The solid black chiral critical  line of the $Z_{2}$ universality differentiating the crossover from the first order region in the $m_{\pi}-m_{K}$ plane at  $\mu=0$,~ends on the strange chiral limit line at the terminal $\pi$ mass $m_{\pi}^{t}\equiv98.15$ MeV in the Fig.~(\ref{fig4a}) for the $m_{\sigma}=400$ MeV.~The transition becomes a smooth crossover everywhere when $m_{\pi}>m_{\pi}^{t}$.~We get smaller $m_{\pi}^{t}\equiv90.03$ MeV in the Fig.~(\ref{fig5a}) when the scalar $\sigma$ meson mass $m_{\sigma}=530$ MeV.~When the  SU(3) symmetric $m_{\pi}=m_{K}$ chiral limit line in the green dash,~intersects the solid black Z(2) chiral critical line,~one gets the critical pion mass $m_{\pi}^{c}$ where the boundary of first order region ends in the $m_{\pi}-m_{K}$ plane at $\mu=0$.~In the QMVT model,~we are finding the  $m_{\pi}^{c}=78.45$ MeV for the $m_{\sigma}=400$ MeV in the Fig.~(\ref{fig4a}) and the  $m_{\pi}^{c}=72.12$ MeV for the $m_{\sigma}=530$ MeV in the Fig.~(\ref{fig5a}).~Recall that the FRG study in Ref.\citep{Resch} finds the  $m_{\pi}^{c}=86$ MeV for $m_{\sigma}=530$ MeV,~the s-MFA study in the Ref. \cite{Schaefer:09} ,~finds $m_{\pi}^{c}\equiv150$ MeV for $m_{\sigma}=800$ MeV and it is pointed out in the Ref. \cite{pisarski24} that the chiral matrix model study using the mean field analysis gives $m_{\pi}^{c}\equiv110$ MeV.~Compiling the critical quantities $(m_{K}^{TCP},~m_{\pi}^{t},~m_{\pi}^{c}) \equiv (242.6,~168.39,~134.16)$ MeV if the $m_{\sigma}=400$ MeV and 
$(m_{K}^{TCP},~m_{\pi}^{t},~m_{\pi}^{c}) \equiv (235.7,~160.75,~128.38)$ MeV if the $m_{\sigma}=530$ MeV for the RQM model study  with the $U_A(1)$ anomaly,~as reported very recently in the Ref.~\cite{vkt25},~it is emphasized that the first order regions shown in the $\mu-m_{K}$ and $m_{\pi}-m_{K}$ planes for the RQM model in the Fig.\ref{fig4b} for the $m_{\sigma}=400$ MeV and the Fig.\ref{fig5b} for the $m_{\sigma}=530$ MeV,~are significantly larger than the corresponding regions of the first order transition in the Fig.\ref{fig4a} and the Fig.\ref{fig5a} for the QMVT model where one respectively finds the $(m_{K}^{TCP},~m_{\pi}^{t},~m_{\pi}^{c}) \equiv (137.2,~98.15,~78.45)$ MeV for the 
$m_{\sigma}=400$ MeV and the $(m_{K}^{TCP},~m_{\pi}^{t},~m_{\pi}^{c}) \equiv (124.7,~90.03,~72.12)$ MeV for the $m_{\sigma}=530$ MeV.~It is well known that the increasing mass of the scalar $\sigma$ meson gives rise to a softer and smoother chiral transition.~It is important to note  that when the $m_{\sigma}$ increases from the 400 to 530 MeV,~the extent of the first order region,~decreases by a small difference of the $(\Delta m_{K}^{TCP},~\Delta m_{\pi}^{t},~\Delta m_{\pi}^{c}) \equiv (12.5,~8.12,~6.33)$ MeV in the QMVT model while correspondingly,~one gets a decrease which is smaller in the RQM model as $(\Delta m_{K}^{TCP},~\Delta m_{\pi}^{t},~\Delta m_{\pi}^{c}) \equiv (6.9,~7.64,~5.78)$ MeV.  

\begin{figure*}[htb]
\subfigure[]{
\label{fig6a} 
\begin{minipage}[b]{0.49\textwidth}
			\centering
			\includegraphics[width=\linewidth]{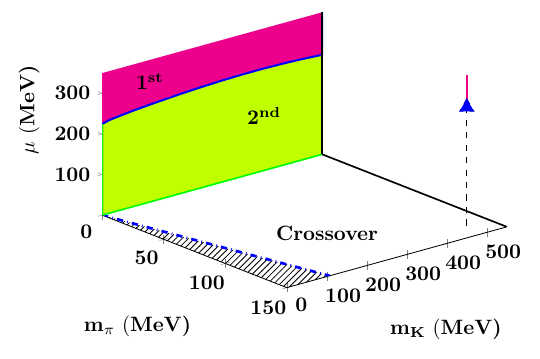}
	\end{minipage}}
	\hfill
	\subfigure[]{
		\label{fig6b} 
		\begin{minipage}[b]{0.49\textwidth}
			\centering 
			\includegraphics[width=\linewidth]{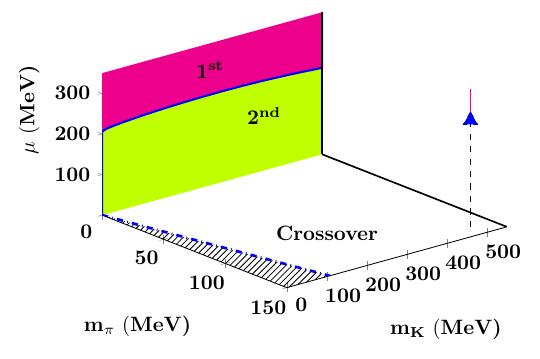}
	\end{minipage}}
\caption{~Columbia plot without the $U_{A}$(1) anomaly is presented in 
Left panel (a) for the QMVT model and right panel (b) for the RQM model when $m_{\sigma}=400$ MeV.~Blue line of tricritical points in the $\mu-m_{K}$ plane for $m_{\pi}=0$,~  differentiates the first order (red) from the second order (light green) chiral transition region.~First order region does not exist in the $m_{\pi}-m_{K}$ plane at $\mu=0$.~The crossover transition vertical line in black dash at the physical point stops at the critical end point with $\mu_{CEP}=291.74 \text{ MeV}$ in (a) for the QMVT model and  $\mu_{CEP}=262.8 \text{ MeV}$ in (b) for the RQM model.}
\label{fig:mini:fig6} 
\end{figure*}

Columbia plot studies are being actively pursued in different approaches in the light of the most recent lattice results.~Here it is relevant to point out that apart from the cubic `t Hooft coupling term,~the sixth order anomalous coupling (which becomes relevant \cite{pisarski24}) term might also govern the change in the size of first order regions as shown in the Fig.1 of a very recent work on the Columbia plot in Ref.\cite{Giacosa}.~They have discussed a scenario where the extent of first order region is maximum when the sixth order anomalous coupling is zero and the first order region decreases corresponding to the increasing sixth order coupling strength while the cubic coupling strength decreases.~They have shown a possibility where phase transition in the whole $m_{\pi}-m_{K}$ plane of the Columbia plot becomes second order,~even though  the $U_{A}(1)$ anomaly is still broken because of the large strength of the  sixth order anomalous coupling term while the cubic coupling strength is zero.~Putting the present work in comparative perspective,~the size of the first-order regions also differ significantly between the two models,~QMVT and RQM in the Fig.\ref{fig:mini:fig4} and Fig.\ref{fig:mini:fig5}.~Note that our work considers only the standard cubic coupling term $c$ which being same as in the QM model,~does not change in the curvature mass based parameter fixing of the QMVT model and significantly reduced first order regions of the Fig.\ref{fig4a} and Fig.\ref{fig5a} are consequence of the very large  smoothing effect of the quark one-loop vacuum fluctuation on the strength of the chiral transition.~In contrast to the above,~the large first order regions in the  Fig.\ref{fig4b} and Fig.\ref{fig5b} of the RQM model Columbia plots are caused by the fact that the $U_{A}(1)$ anomaly strength $c$ which contains a condensate dependent part,~ gets significantly enhanced when the meson self energies due to quark loops are evaluated using the meson pole masses and parameters are fixed on-sell.~The smoothing effect of the quark one-loop vacuum fluctuation is moderate in RQM Model.

The small first order regions in the Fig.\ref{fig4a} and Fig.\ref{fig5a} for the QMVT and
significantly large  first order regions in the Fig.\ref{fig4b} and Fig.\ref{fig5b} for the RQM model,~increase near the origin (chiral limit) of the vertical $\mu-m_{K}$ plane at the $m_{\pi}=0$ when the chemical potential $\mu$ is increased.~In all the above model scenarios,~the chiral critical surface has a positive curvature \cite{forcrd2}.~Therefore,~a critical end point exists in the Fig.\ref{fig4a} and  Fig.\ref{fig5a} ( Fig.\ref{fig4b}  and Fig.\ref{fig5b}) for the QMVT model (RQM model) and gets marked by the solid blue arrow in the $\mu-T$ plane of figures where the dashed black vertical line,~of the crossover transition at the physical  point,~intersects the chiral critical surface.~The chiral tricritical solid blue line has a large positive slope initially when it begins from the $m_{K}^{TCP}=137.2 \text{ MeV at } \mu=0$ MeV in the QMVT model in Fig.~\ref{fig4a} for $m_{\sigma}=400$ MeV.~The above slope decreases successively for the larger $\mu$ and $ m_{K}$.~The average slope of tricritical line is only .078 between the points $(m_{K},\mu)=(260,223.26)$ MeV and $(m_{K},\mu)=(500,242.05)$ MeV  and it becomes  zero for the $m_{K}=400-500$ MeV range as $\mu=241.98$ MeV for the $m_{K}=400$ MeV,~becomes only $\mu=242.05$ MeV for the $m_{K}=500$ MeV.~When $m_{K}=550$ MeV,~$\mu$ decreases slightly to $\mu=239.72$ MeV and the slope becomes slightly negative.~Thus the tricritical line in the Fig.~\ref{fig4a} shows a complete saturation between the $m_{K}=400-500$ MeV as it becomes parallel to the $m_{K}$ axis (for $m_{\pi}=0$) with zero slope.~In contrast to the QMVT model,~the RQM model tricritical line for $m_{\sigma}=400$ MeV,~starting from the $m_{K}^{TCP}=242.6, \ \mu=0$ MeV in the Fig.~\ref{fig4b},~shows a near saturation trend (not a complete saturation) for chemical potentials which are about 52 MeV smaller than the $\mu$ values in the QMVT model around which the saturation occurs.~In the RQM model,~the $\mu=184.7$ MeV at $m_{K}=450$ MeV becomes $\mu=188.25$ MeV at $m_{K}=500$ MeV with a .071 slope of the tricritical line in the Fig.~\ref{fig4b} and this slope becomes smaller as it is .049 when the $\mu=188.25$ MeV at $m_{K}=500$ MeV becomes $\mu=190.7$ MeV at $m_{K}=550$ MeV.

\begin{figure*}[htb]
\subfigure[]{
\label{fig7a} 
\begin{minipage}[b]{0.49\textwidth}
			\centering
			\includegraphics[width=\linewidth]{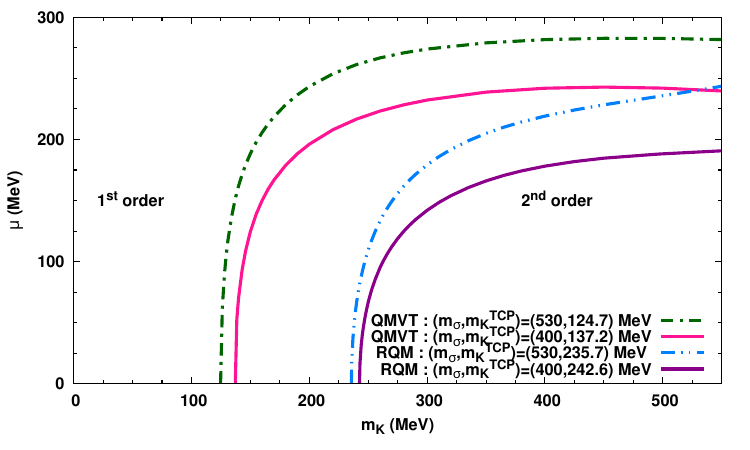}
	\end{minipage}}
	\hfill
	\subfigure[ ]{
		\label{fig7b} 
		\begin{minipage}[b]{0.49\textwidth}
			\centering 
			\includegraphics[width=\linewidth]{ 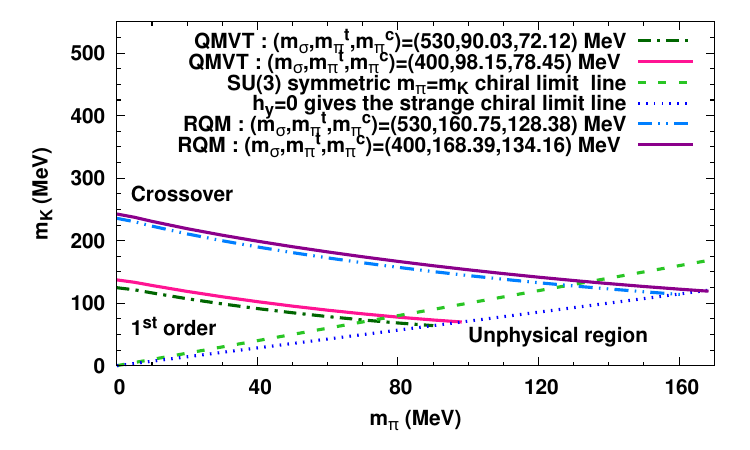}
	\end{minipage}}
	\caption{Left panel (a) (Right panel (b)) shows the comparison of lines and regions in the vertical $\mu-m_{K}$, $m_{\pi}=0$ (horizontal $m_{\pi}-m_{K}$, $\mu=0$) planes of the three dimensional QMVT and RQM model Columbia plots for the $m_{\sigma}=400, \ 530$ MeV  in the Fig.(\ref{fig4a}),~Fig.(\ref{fig5a}) and Fig. (\ref{fig4b}),~Fig.(\ref{fig5b}).~QMVT and  RQM model tricritical lines for the $m_{\sigma}=400 \text{ and } 530$ MeV,~as marked and explained in the Fig.(a),~start from the $m_{K}^{TCP}$ on the $m_{K}$ axis and demarcate the first order region above them on their left hand side  from the second order region below them on their right hand side.~QMVT and RQM model Z(2) critical lines for the $m_{\sigma}=400 \text{ and } 530$ MeV,~as marked and explained in the Fig.(b), separating the crossover transition region above them from the first order region that lies below,~intersect the SU(3) symmetric $m_{\pi}=m_{K}$ chiral limit line at the critical pion mass $m_{\pi}^{c}$ and terminate on the strange chiral limit line satisfying  the $h_{y}=0=h_{y0}$,~at the terminal pion mass $m_{\pi}^{t}$.}
\label{fig:mini:fig7}
\end{figure*}

The QMVT model tricritical line for the $m_{\sigma}=530$ MeV that starts from the $m_{K}^{TCP}=124.7, \text{ at } \mu=0$ MeV in the Fig.~\ref{fig5a},~shows a good saturation trend similar to the $m_{\sigma}=400$ MeV case when chemical potential ($\mu \equiv 283$ MeV) is significantly higher.~The average slope of the above line is only .066 between the points $(m_{K},\mu)=(260,266.82)$ MeV and $(m_{K},\mu)=(500,282.77)$ MeV and the slope becomes  zero for the $m_{K}=450-500$ MeV range as $\mu=282.85$ MeV for the $m_{K}=450$ MeV,~becomes only $\mu=282.77$ MeV for the $m_{K}=500$ MeV.~When $m_{K}=550$ MeV,~$\mu$ decreases slightly to $\mu=281.81$ MeV and the slope becomes slightly negative.~In complete contrast to the QMVT model,~the tricritical line of RQM model for $m_{\sigma}=530$ MeV starting from the $m_{K}^{TCP}=235.7, \text{ at } \mu=0$ MeV  in Fig.~\ref{fig5b},~does not show a near saturation trend like the one that we get for the $m_{\sigma}=400$ MeV case.~The slope 0.1852 between the points ($\mu=219.14,m_{K}=400.0$) MeV and  ($\mu=228.4,m_{K}=450.0$) MeV on the tricritical line in the Fig.~\ref{fig5b},~decreases to the 0.1478 when one goes from the point ($\mu=228.4,m_{K}=450.0$) MeV to the point ($\mu=235.79,m_{K}=500.0$) MeV but it increases and becomes 0.1566 between the point ($\mu=235.79,m_{K}=500.0$) MeV and ($\mu=243.62,m_{K}=550.0$) MeV.~The critical end point marked by the solid blue arrow,~gets located at the $\mu_{CEP}=288.03$ MeV for the $m_{\sigma}=400$ MeV and the $\mu_{CEP}=322.16$ MeV for the $m_{\sigma}=530$ MeV case in the QMVT model while in comparison the position of the critical end point is at the $\mu_{CEP}=243.29$ MeV for the $m_{\sigma}=400$ MeV and the $\mu_{CEP}=275.96$ MeV for the $m_{\sigma}=530$ MeV in the RQM model.

In the absence of $U_A(1)$ anomaly,~no first order region is found in the $m_{\pi}-m_{K}$ plane at $\mu=0$ either in the Fig.~(\ref{fig6a}) for the QMVT model or in the Fig.~(\ref{fig6b}) for the RQM model when $m_{\sigma}=400$ MeV.~The above result is similar to the findings of the e-MFA:QM model FRG study in the Ref. \cite{Resch} and it is consequence of the effect of quark one-loop vacuum fluctuation.~For the  chemical potentials $\mu<\mu_{c}$,~whatever value  the kaon mass $m_{K}$ takes in the light chiral limit $m_{\pi}=0$,~one finds a second order chiral transition everywhere in the $\mu-m_{K}$ plane of the Fig.~(\ref{fig6a}) and Fig.~(\ref{fig6b}) respectively 
for the QMVT  and  RQM model.~The chemical potential for which the first order transition 
reappears in the Fig.~(\ref{fig6a}) for the QMVT model,~is very high  $\mu_{c}=225.4$ MeV.~The tricritical line,~demarcating the first order transition region above it from the second order chiral transition region below,~has a small positive slope until the $(\mu_{c},m_{K})=(248.2,250)$ MeV which becomes nearly constant in the range $m_{K}=350-450$ MeV after the $\mu_{c}$ attains maximum at 251.2 MeV when the $m_{K}=350$.~Afterwards,~the slope decreases very slowly when the $\mu_{c}$ becomes 245.32 MeV at the $m_{K}=550$ MeV.~The RQM model $\mu-m_{K}$ plane at  $m_{\pi}=0$ in the Fig.~(\ref{fig6b}) has similar features as above but here,~the first order (second order) region  above (below) the tricritical line,~is larger (smaller) than the corresponding  first order (second order) region seen in the Fig.~(\ref{fig6a}) for the QMVT model.~The first order region in the RQM model appears again for the $\mu_{c}=204.6$ MeV which is smaller than the corresponding QMVT model  $\mu_{c}=225.4$ MeV and the RQM model tricritical line has a small positive slope until $\mu_{c}=223.2,m_{K}=240$ MeV which becomes constant for the range $m_{K}=240-350$ MeV and then decreases very slowly when the $\mu_{c}$ becomes 213.2 MeV at the $m_{K}=550$ MeV.~The critical end point  corresponding to  the physical point in the  Fig.~(\ref{fig6a}) of the QMVT model,~is marked by the solid blue arrow at the $\mu_{CEP}=291.74$ MeV while the RQM model critical end point is pointed by the solid blue arrow at the $\mu_{cep}=262.8$ MeV in the Fig.~(\ref{fig6b}).


The lines and regions,~in the vertical $\mu-m_{K}$ plane (at $m_{\pi}=0$) of the three dimensional Columbia plots of the QMVT (RQM) model  in the Fig.~(\ref{fig4a}) (Fig.~(\ref{fig4b})) for the $m_{\sigma}=400$ MeV and Fig.~(\ref{fig5a}) (Fig.~(\ref{fig5b})) for the $m_{\sigma}=530$ MeV,~have been replotted in the Fig.~(\ref{fig7a}) to see the comparison in one figure.~The blue solid line of the tricritical points in the vertical $\mu-m_{K}$ planes of the above figures,~has been replotted respectively for the QMVT and RQM model,~by the dash dotted deep green line starting from the $m_{K}^{TCP}=124.7$ MeV  (solid red line originating from the $m_{K}^{TCP}=135.2$ MeV)  for the $m_{\sigma}=530 (400)$ MeV and the dash double dotted sky blue line beginning from the $m_{K}^{TCP}=235.7$ MeV (solid indigo line originating from the $m_{K}^{TCP}=242.6$ MeV) for the $m_{\sigma}=530(400)$ MeV.~It is amply clear from the Fig.~(\ref{fig7a}) that the first order region in the QMVT model is  significantly smaller  than the corresponding first order region in the RQM model.~The QMVT model tricritical line becomes parallel to the $m_{K}$ axis  showing good saturation trend both for the $m_{\sigma}=400 \text{ and } 530$ MeV while the RQM model tricritical line shows near saturation like trend for $m_{\sigma}=400$ MeV.~The saturation trend gets spoiled in the RQM model tricritical line for the $m_{\sigma}=530$ MeV whose slope starts increasing for the $m_{K}>500$ MeV.~The  saturation trend in the QMVT model sets up for about 47-52 MeV higher chemical potential than the value of the chemical potential around which   saturation like trend sets up in the RQM Model.

\begin{table*}[!htbp] 
\caption{This table presents the  quantities $m_{K}^{TCP},\ m_{s}^{TCP}, \ m_{\pi}^{t}, \ m_{ud}^{t}, \ m_{\pi}^{c},\ m_{ud}^{c}$ and the $\beta_{cep}:m_{\pi,cep}^*,m_{K,cep}^* $ calculated for $m_{\sigma}=400 \text{ and } 500$ MeV in the QMVT and RQM model.~RQM model results for $m_{\sigma}=530$ MeV obtained using the inputs of large $N_c$ standard U(3) ChPT  termed as input set $ \text{M}_{\eta}$-II in Ref.\cite{vkt25},~is also presented.~Some of these quantities as obtained  in the e-MFA:QM model FRG study for $m_{\sigma}=530$ MeV and the QM model s-MFA study for $m_{\sigma}=800$ MeV,~are also presented for comparison.~The second order chiral transition and the TCP is absent on the $m_\pi=0$ axis in the QM model s-MFA study.~The last two rows show the critical pion (light quark) mass $m_{\pi}^{c}$ ($m_{ud}^{c}$) for three degenerate flavors at $\mu=0$ from different LQCD studies whose results are labeled with the corresponding fermion implementation and the number   of time slices $N_t$.}
	\label{tab:table3}
	\begin{tabular}{p{0.169\textwidth}| p{0.085\textwidth} |p{0.08\textwidth}| p{0.06\textwidth}|p{0.06\textwidth}| p{0.08\textwidth}|p{0.08\textwidth} |p{0.08\textwidth}| p{0.085\textwidth} | p{0.085\textwidth}| p{0.085\textwidth}}
		\hline 
		Model&$m_{\sigma}$&$m_{K}^{TCP}$&$m_{s}^{TCP}$&$m_{\pi}^{t}$ &$m_{ud}^{t}$ &$m_{\pi}^{c}$&$m_{ud}^{c}$&$\beta_{cep}$& $m_{\pi,cep}^{*} $ & $m_{K,cep}^{*}$\\
		\hline 
QMVT&$400$&137.2& 8.06 &98.15& 2.05 &78.45&1.31 &0.2215 & 30.57& 109.88\\
		\hline
RQM&$400$&242.6& 25.53 &168.39& 6.13 &134.16&3.84 &0.3795 & 52.38& 188.25\\
		\hline
QMVT&$530$&124.7&6.64&90.03&1.72 & 72.12&1.1&0.2028 &27.98 &100.58\\
		\hline		
RQM&$530$&235.7&24.08&160.75&5.58 & 128.38&3.51&0.3642 &50.26 &180.64\\
		\hline
RQM : $\text{M}_{\eta}$-II \cite{vkt25}&$530$&239.58& 23.47 &161.35&5.44 & 128.04&3.45 &0.36887 &50.9 &182.96\\
		\hline
		e-MFA:QM-FRG\cite{Resch}&$530$&169&- & 110&3.0 &86 &-& 0.20 & 27.6 &99.2\\
		\hline
		s-MFA:QM\cite{Schaefer:09}&$800$&No TCP&--& 177&- &150 &- & 0.488 & 67.34 &242.05\\
		\hline %
	\end{tabular}
\begin{tabular}{p{0.05\textwidth}| p{0.135\textwidth} |p{0.11\textwidth}| p{0.135\textwidth}|p{0.11\textwidth}| p{0.11\textwidth}|p{0.11\textwidth} |p{0.11\textwidth}| p{0.11\textwidth}}   
LQCD Study&$N_{t}=4$,Standard Staggered $m_{\pi}^c$ &$N_{t}=4$, p4 Staggered $m_{\pi}^c$&$N_{t}=6$,Standard Staggered $m_{\pi}^c$&$N_{t}=6$,Stout Staggered $m_{\pi}^c$&Wilson Clover $N_{t}=6$, $m_{\pi}^c$&$N_{t}=6$, HISQ Staggered $m_{\pi}^c$&Wilson Clover $N_{t}=8$, $m_{\pi}^c$& $m_{ud}^c$ : Möbius Domain Wall \\
\hline
Ref.&$\sim$290\ \cite{karsch2}&$\sim$67\ \cite{karsch3}&$\sim$150 \ \cite{forcrd}&$\sim$0 \ \cite{varnho}&$\sim$300\ \cite{jin}&$\le$50 \ \cite{Bazav}&$\le$170\ \cite{jin2}&$\le$4 \ \cite{zhang24}\\
\hline 
\end{tabular}
\end{table*}
~The Fig.~(\ref{fig7b}) depicts,~the comparison of the lines and regions in the horizontal $m_{\pi}-m_{K}$ plane (at $\mu=0$) of the three dimensional Columbia plots of the QMVT (RQM) model  in the Fig.~(\ref{fig4a}) (Fig.~(\ref{fig4b})) for the $m_{\sigma}=400$ MeV and Fig.~(\ref{fig5a}) (Fig.~(\ref{fig5b})) for the $m_{\sigma}=530$ MeV.~The Z(2) critical black solid line,~in the horizontal $\mu-m_{K}$ plane of the above figures,~has been replotted,~by the dash dotted deep green line (solid red line )  for the $m_{\sigma}=530 (400)$ MeV  in the QMVT  model and the dash double dotted sky blue line (solid indigo line) for the $m_{\sigma}=(530)400$ MeV in the RQM model.~It is recalled that the QMVT and RQM model Z(2) critical lines for the $m_{\sigma}=400 (530)$ MeV,~intersect the SU(3) symmetric $m_{\pi}=m_{K}$ chiral limit line at the critical pion mass $m_{\pi}^{c}$ and terminate on the strange chiral limit line satisfying  the $h_{y}=0=h_{y0}$,~at the terminal pion mass $m_{\pi}^{t}$.~One finds the $(m_{\pi}^{t},m_{\pi}^{c})=(98.15,78.45)$($(m_{\pi}^{t},m_{\pi}^{c})=(90.03,72.12)$) MeV in the QMVT model while in the RQM model one gets the $(m_{\pi}^{t},m_{\pi}^{c})=(168.39,134.16)$ ($(m_{\pi}^{t},m_{\pi}^{c})=(160.75,128.38)$) MeV for the $m_{\sigma}=400 (530)$ MeV.~The Fig.~(\ref{fig7b}) clearly demonstrates that the first order area underneath the crossover region of the QMVT model,~is  significantly smaller  than the corresponding first order area lying below the crossover region of the RQM model.~Table (\ref{tab:table3}) shows the comparison of our present QMVT model results with those of the e-MFA:QM model FRG  work \cite{Resch} and RQM model work reported very recently in the Ref.~\cite{vkt25}.~The  FRG work reports the $(m_{K}^{TCP},~m_{\pi}^{t},~m_{\pi}^{c}) = (169,~110,~86)$ MeV for the $m_{\sigma}=530$ MeV,~and we are finding $(m_{K}^{TCP},~m_{\pi}^{t},~m_{\pi}^{c}) = (124.7,~90.03,~72.12)$ MeV in the present QMVT model work while one notes the $(m_{K}^{TCP},~m_{\pi}^{t},~m_{\pi}^{c}) \equiv (235.7,~160.75,~128.38)$ MeV in the RQM model work.~The critical quantities,~computed in the RQM model with the large $N_{c}$ standard U(3) ChPT input set $\text{M}_{\eta}$-II when the $m_{\sigma}=530$ MeV,~are slightly different as the $(m_{K}^{TCP},~m_{\pi}^{t},~m_{\pi}^{c}) \equiv (239.58,~161.35,~128.04)$ MeV~\cite{vkt25}.~While being much smaller than their numerical values computed in the RQM model,~the critical quantities $(m_{K}^{TCP},~m_{\pi}^{t},~m_{\pi}^{c})$ found in the present QMVT model work,~are somewhat smaller but stand closer to the corresponding values reported in the e-MFA QM model FRG study of the Ref.\cite{Resch}.~The tricritical lines in the $\mu-m_{K}$ plane,~show similar saturation trends for the QMVT model and the e-MFA:QM model FRG study while the saturation trend for the RQM model tricritical line gets spoiled.

~The first order regions in the $\mu-m_{K}$ and $m_{\pi}-m_{K}$ planes of the QMVT model Columbia plot,~ while being much smaller than the corresponding first order regions for the RQM model,~are similar but moderately smaller than the extent of the first order regions computed for the e-MFA:QM model FRG study.~Here,~it is relevant to point out that the $\sigma$ mass,~in the e-MFA:QM model FRG  study,~is calculated from the curvature of the effective potential that gets evaluated by the FRG flow equation.~Recall that the curvature masses of mesons are used for fixing the parameters in the QMVT model and this context of incorporating quark one-loop vacuum correction in the QMVT model is analogous to that of the e-MFA:QM model FRG study in the Ref.\cite{Resch}.~It is important to point out that even though the fraction $\beta_{cep}$,~by which the  pion and kaon masses are to be reduced in the $m_{\pi}-m_{K}$ plane such that the chiral crossover ends into a second order critical end point giving the first order transition region afterwards when $\beta<\beta_{cep}$,~is almost the same in both the models as $(\beta_{cep}|_{m_{\sigma}=530\text{MeV}:\text{QMVT}}=0.2028)\simeq(\beta_{cep}|_{m_{\sigma}=530\text{MeV}:\text{eMFA;QM-FRG}}\equiv0.20)$,~the extent first order transition regions in the QMVT model Columbia plot,~is somewhat smaller than what is found in the  e-MFA:QM model Columbia plot as the critical quantities $(m_{K}^{TCP},~m_{\pi}^{t},~m_{\pi}^{c})|_{\text{QMVT}} = (124.7,~90.03,~72.12) < (m_{K}^{TCP},~m_{\pi}^{t},~m_{\pi}^{c})|_{\text{eMFA:QM-FRG}} = (169,~110,~86)$ MeV for the $m_{\sigma}=530$ MeV.~The possible reason of the above difference lies in the different methods adopted for performing chiral limit study.~Recall that the $\sigma $ curvature mass  $m_{\sigma}=530$ MeV does not change towards the chiral limit in the QMVT model when the ($m_{\pi},m_{K}$) dependent ChPT scaling relations for the $f_{\pi},f_{K} \text{ and } M_{\eta}$,~are used for calculating the change in model parameters away from the physical point whereas in contrast,~the vacuum value of the $m_{\sigma}$ decreases significantly towards the chiral limit in the e-MFA:QM model work where the $f_{\pi}, M_{\eta}$ and the model parameters are kept fixed as at the physical point when the scale $\Lambda$  of the initial effective action for the FRG flow is successively increased to $\Lambda=700,1000,1100 \text{ and } 1143$ for compensating the  successive decrease in the chiral symmetry breaking source strengths $j_{x} \text{ and } j_{y}$ by the fractions $\alpha=\sqrt{\beta}=1.0,0.17,0.04 \text{ and } 0$.~The quark one-loop vacuum fluctuation  generates quite a large smoothing effect on the strength of chiral transition in the e-MFA:QM model FRG study and this softening of the chiral transition by the fluctuation is very large (larger than that of the e-MFA:QM FRG study) in the QMVT model.~Note that the full FRG QM model study where meson thermal and vacuum quantum fluctuations are also included,~finds a very small $m_{\pi}^{c}\equiv17$ MeV for which Pisarski et. al. \cite{pisarski24} has observed that this is partly because the approximation used in Ref. \cite{Resch} is known to overestimate the masonic fluctuations that tend to soften the chiral transition \cite{PawlRen}.~Note also that  the Ref. \cite{fejoHastuda} has cautioned that the LPA in the FRG studies,~completely neglects the wave function renormalization which might affect the final result.~Recall that the quark one-loop vacuum  fluctuation softens the chiral transition moderately in the RQM model because here the on-shell renormalization of parameters gives rise to stronger $U_{A}(1)$ anomaly strength $c$ together with a weaker light and strange chiral symmetry breaking strengths $h_{x}\text{ and } h_{y}$.~Hence one gets significantly large first order regions in the RQM model Columbia plots \cite{vkt25}.

\subsection{\bf{Comparing  the light-strange quark mass $\bf {m_{ud}-m_{s}}$ and $\bf {\mu-m_{s}}$ planes of the QMVT and RQM model Columbia plots for two different values of $\bf {m_{\sigma}}$ }}
\label{subsec:Colpqm}

\begin{figure*}[htb]
\subfigure[]{
\label{fig8a} 
\begin{minipage}[b]{0.49\textwidth}
			\centering
			\includegraphics[width=\linewidth]{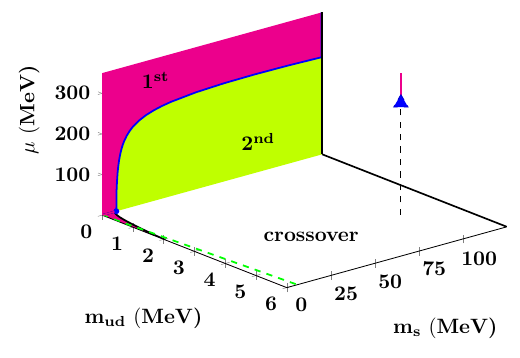}
	\end{minipage}}
	\hfill
	\subfigure[]{
		\label{fig8b} 
		\begin{minipage}[b]{0.49\textwidth}
			\centering 
			\includegraphics[width=\linewidth]{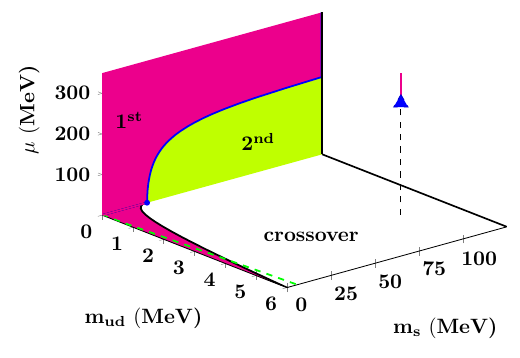}
	\end{minipage}}
	\caption{~Columbia plot with the $U_{A}$(1) anomaly in the light-strange quark mass and chemical potential  ($m_{ud}-m_{s} \text{ and } \mu$) plane is presented in the left (right) panel (a) ((b)) for the QMVT (RQM) model when $m_{\sigma}=400$ MeV.~The $\mu-m_{s}$ ($m_{ud}-m_{s}$) plane presents the chiral transition for the $m_{ud}=0$ ($\mu=0$).~The second order $Z(2)$ critical solid black line demarcating the crossover from the first order region,~intersects the SU(3) symmetric $m_{ud}=m_{s}$ green dash line respectively at the critical light quark mass of $m_{ud}^{c}=1.31$ and $m_{ud}^{c}=3.84$ MeV in (a) for the QMVT model and (b) for the RQM model.~It terminates at the terminal light quark mass of $m_{ud}^{t}=2.05$ MeV in (a) for the QMVT model and $m_{ud}^{t}=6.13$ MeV in (b) for the RQM model.~Solid blue line of the tricritical points that separates the second  and first order regions in the $\mu-m_{s}$ plane,~starts on the $m_{ud}=0$ axis from the blue dot respectively at the $m_{s}^{TCP}=8.06$ MeV in (a) for the QMVT model and $m_{s}^{TCP}=25.53$ MeV in (b) for the RQM model.~The physical point $(m_{\pi},m_{K})=(138,496)$ MeV is at the $(m_{ud},m_{s})=(4.0,99.77)$ MeV in the $m_{ud}-m_{s}$ plane where the black dash vertical crossover line  ends at the critical end point with $\mu_{CEP}=288.03 \text{ MeV}$ in (a) for the QMVT model and  $\mu_{CEP}=243.29 \text{ MeV}$ in (b) for the RQM model.~The vertical red line shows the first order transition.}
	\label{fig:mini:fig8} 
\end{figure*}

\begin{figure*}[htb]
\subfigure[]{
\label{fig9a} 
\begin{minipage}[b]{0.49\textwidth}
			\centering
			\includegraphics[width=\linewidth]{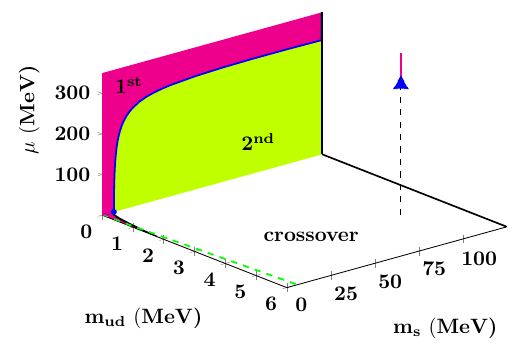}
	\end{minipage}}
	\hfill
	\subfigure[]{
		\label{fig9b} 
		\begin{minipage}[b]{0.49\textwidth}
			\centering 
			\includegraphics[width=\linewidth]{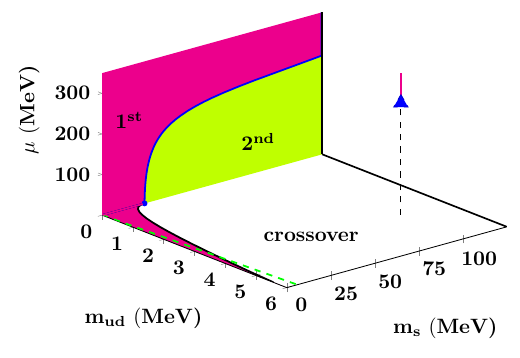}
	\end{minipage}}
\caption{~Left panel (a) (right panel (b)),~depicts the Columbia plot in the light-strange quark mass and chemical potential  ($m_{ud}-m_{s} \text{ and } \mu$) plane of the QMVT (RQM) model when $U_{A}$(1) anomaly is present and the $m_{\sigma}=530$ MeV.~The $m_{ud}=0$ on the $\mu-m_{s}$ plane while $\mu=0$ for the $m_{ud}-m_{s}$ plane.~The critical points,~lines,~first, second order,~crossover regions and other features are defined similar to the Fig.(\ref{fig:mini:fig8}).~The left panel (a) Columbia plot,~gives the $(m_{s}^{TCP},~m_{ud}^{t},~m_{ud}^{c}) \equiv (6.64,~1.72,~1.10)$ MeV for the QMVT model.~For the RQM Model,~Columbia plot in (b)  gives $(m_{s}^{TCP},~m_{ud}^{t},~m_{ud}^{c}) \equiv (24.08,~5.58,~3.51)$ MeV.~Vertical line of crossover transition at the physical point, terminates at the critical end point with $\mu_{CEP}=322.16 \text{ MeV}$ in (a) for the QMVT model and  $\mu_{CEP}=275.95 \text{ MeV}$ in (b) for the RQM model.}
	\label{fig:mini:fig9} 
\end{figure*}

In order to make the results of effective model studies available for comparing and contrasting with the LQCD results and studies in other approaches,~the Columbia plots in the $m_{\pi}-m_{K} \text{ and } \mu-m_{K}$ plane,~are required to be redrawn in the light-strange quark mass and chemical potential $m_{ud}-m_{s} \text{ and } \mu-m_{s} $ planes.~Using  the Eqs.~(\ref{Aqpim}) and (\ref{qmr}) given at the end of the 
section~\ref{subsec:Chpt} and fixing the light quark mass  $m_{ud}=4$ MeV,~the QMVT model and RQM model Columbia plots in the $m_{\pi}-m_{K}$ plane for the $m_{\sigma}=400 \text{ and } 530$ MeV,~are mapped below into the corresponding  light-strange quark mass $m_{ud}-m_{s}$ plane.~The quark mass $m_{ud}-m_{s}$ plane and the $\mu-m_{s}$ plane of Columbia plot when the $m_{\sigma}=400$ MeV,~are presented in the Fig.~(\ref{fig8a}) for the QMVT model and the Fig.~(\ref{fig8b}) for the RQM model.~The chiral transition for the light chiral limit $m_{ud}=$0,~is shown in the $\mu-m_{s}$ plane and the  $m_{ud}-m_{s}$ plane shows the chiral transition at zero chemical potential $\mu=$0.

The tricritical solid blue line made by the tricritical points in the $\mu-m_{s}$ plane at
$m_{ud}=0$ when the $m_{\sigma}=400$ MeV,~starts from the $m_{s}^{TCP}=8.06$ MeV
and $m_{s}^{TCP}=25.53$ MeV respectively in  the Fig.~(\ref{fig8a}) for the QMVT model and  Fig.~(\ref{fig8b}) for the RQM model.~The solid black second order $Z(2)$ critical line in the $m_{ud}-m_{s}$ plane at $\mu=0$,~intersects the $SU(3)$ symmetric $m_{ud}=m_{s}$ green dash line respectively in the Fig.~(\ref{fig8a}) and Fig.~(\ref{fig8b}) at the critical light quark mass of $m_{ud}^{c}=1.31$ MeV for the QMVT model and the $m_{ud}^{c}=3.84$ MeV for the RQM model.~Furthermore this line terminates on the $m_{ud}$ axis (for $m_{s}=0$) respectively at the terminal light quark mass of,~$m_{ud}^{t}=2.05$ MeV in the Fig.~(\ref{fig8a}) for QMVT model and $m_{ud}^{t}=6.13$ MeV in the Fig.~(\ref{fig8b}) for RQM model.~The black dashed vertical line denoting the crossover transition at the physical point,~ends in the solid blue triangle depicting the critical end point respectively at $\mu_{CEP}=288.09$ MeV in the Fig.~(\ref{fig8a}) for QMVT model and 			$\mu_{CEP}=243.29$ MeV in the Fig.~(\ref{fig8b}) for RQM model.~The different critical lines,~crossover,~first order and second order transition regions and critical quantities in the Fig.~(\ref{fig9a}) and (\ref{fig9b}) respectively for the QMVT and RQM model when the $m_{\sigma}=530$ MeV,~are defined exactly like those in the
Fig.~(\ref{fig8a}) and (\ref{fig8b}) for the $m_{\sigma}=400$ MeV case.~One finds the  $(m_{s}^{TCP},m_{ud}^{t},m_{ud}^{c})=(6.64,1.72,1.1)$ MeV for the $( \ m_{K}^{TCP},m_{\pi}^{t},m_{\pi}^{c})=(124.7,90.03,72.12)$ MeV in the QMVT model and  $(m_{s}^{TCP},m_{ud}^{t},m_{ud}^{c})=(24.08,5.58,3.51)$ MeV corresponding to the $( \ m_{K}^{TCP},m_{\pi}^{t},m_{\pi}^{c})=(235.7,160.75,128.38)$ MeV in the RQM model when the $m_{\sigma}=530$ MeV.~It is worth emphasizing that similar to the above mentioned pattern of numerical values of the QMVT and RQM model critical quantities when the $m_{\sigma}=530$ MeV,~the critical quantities  $(m_{s}^{TCP},m_{ud}^{t},m_{ud}^{c})=(8.06,2.05,1.31)$ MeV corresponding to the $( \ m_{K}^{TCP},m_{\pi}^{t},m_{\pi}^{c})=(137.2,98.15,78.45)$ MeV in the QMVT model for the $m_{\sigma}=400$ MeV,~are significantly smaller than the numerical values $(m_{s}^{TCP},m_{ud}^{t},m_{ud}^{c})=(25.53,6.13,3.84)$ MeV corresponding to the  $( \ m_{K}^{TCP},m_{\pi}^{t},m_{\pi}^{c})=(242.6,168.39,134.16)$ MeV of these critical quantities in the RQM model.

\begin{figure*}[htb]
\subfigure[]{
\label{fig10a} 
\begin{minipage}[b]{0.49\textwidth}
			\centering
			\includegraphics[width=\linewidth]{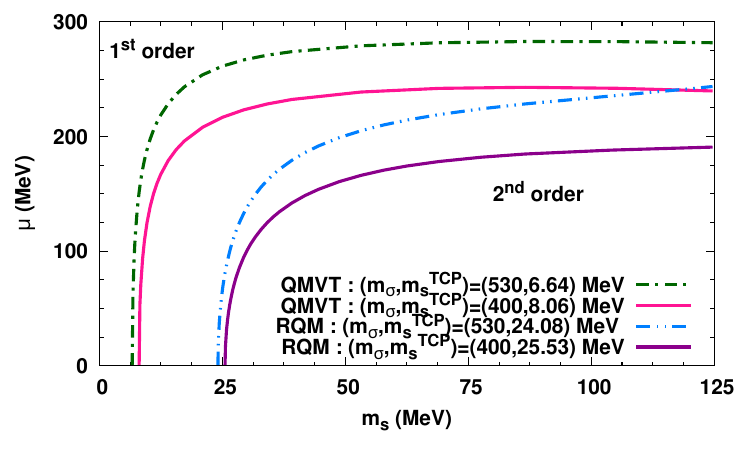}
	\end{minipage}}
	\hfill
	\subfigure[]{
		\label{fig10b} 
		\begin{minipage}[b]{0.49\textwidth}
			\centering 
			\includegraphics[width=\linewidth]{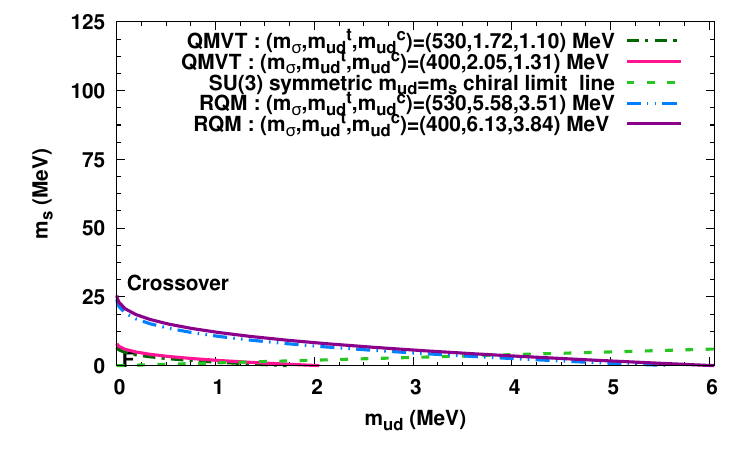}
	\end{minipage}}
	\caption{Left panel (a) (Right panel (b)) shows the comparison of lines and regions in the vertical $\mu-m_{s}$, $m_{ud}=0$ (horizontal $m_{ud}-m_{s}$, $\mu=0$) planes of the three dimensional QMVT and RQM model Columbia plots for the $m_{\sigma}=400, \ 530$ MeV  in Fig.(\ref{fig8a}),~Fig.(\ref{fig9a}) and Fig. (\ref{fig8b}),~Fig.(\ref{fig9b}).~QMVT and RQM model tricritical lines for the $m_{\sigma}=400 \text{ and } 530$ MeV,~as marked and explained in Fig.(a),~start from the $m_{s}^{TCP}$ on the $m_{s}$ axis and demarcate the first order region above them on their left hand side  from the second order region below them on their right hand side.~QMVT and RQM model Z(2) critical lines for the $m_{\sigma}=400 \text{ and } 530$ MeV,~as marked and explained in Fig.(b), separating the crossover transition region above them from the first order region below them,~intersect the SU(3) symmetric $m_{ud}-m_{s}$ chiral limit line at the critical light quark mass $m_{ud}^{c}$ and terminate ~at the terminal light quark mass $m_{ud}^{t}$.}
	\label{fig:mini:fig10} 
\end{figure*}

The lines and regions of the vertical and horizontal planes of the QMVT and RQM model three dimensional Columbia plots respectively in the Fig.~(\ref{fig8a}) and Fig.~(\ref{fig8b}) for the $m_{\sigma}=400$ MeV as well as  the Fig.~(\ref{fig9a}) and   Fig.~(\ref{fig9b} )for the $m_{\sigma}=530$ MeV,~have been replotted respectively in the 
two dimensional $\mu-m_{s}$ plane at $m_{ud}=0$ in the Fig.~(\ref{fig10a}) and $m_{ud}-m_{s}$ plane at $\mu=0$ in the Fig.~(\ref{fig10b}),~to get a comprehensive picture of the  details of comparison in one figure.~The plot legends have marked and figure caption has  explained the tricritical lines tagged with the  $(m_{\sigma},m_{s}^{TCP})$ values and the Z(2) critical lines tagged with  the $(m_{\sigma},m_{\pi}^{t},m_{\pi}^{c})$ values respectively in the Fig.~(\ref{fig10a}) and Fig.~(\ref{fig10b}) for the QMVT as well as the RQM model.~The Fig.~(\ref{fig10a}) and Fig.~(\ref{fig10b}) lead us to the conclusion that the first order region in the QMVT model is  much smaller  than the corresponding first order region in the RQM model.~Becoming parallel to the $m_{s}$ axis for  $m_{s}>69$ MeV,~the tricritical lines,~show good saturation in the QMVT model whereas the  near saturation like trend of the RQM model tricritical line when $m_{\sigma}=400$ MeV,~gets spoiled for the $m_{\sigma}=530$ MeV  tricritical line whose slope starts increasing for the $m_{s}>105$ MeV.~It is important to note  that the critical light quark mass obtained in the QMVT model study is  $m_{ud}^c=1.31 \text{ and } 1.1$ MeV respectively for the $m_{\sigma}=400 \text{ and } 530$ MeV while in the RQM model study,~one finds the  $m_{ud}^c=3.84 \text{ and } 3.51$ MeV respectively for the $m_{\sigma}=400 \text{ and } 530$ MeV. ~The recent  LQCD study in Ref.~\cite{zhang24} using  Möbius domain wall fermions  finds that $m_{ud}^c\le$4 MeV.~Results from different LQCD studies for the critical pion mass $m_{\pi}^c$ and the critical light quark mass $m_{ud}^c$ are compiled in the Table(\ref{tab:table3}) for ready reference.

\section{Summary}
\label{secIV}
The consistent chiral limit has been explored in the QMVT model \cite{schafwag12,chatmoh1,vkkt13,Rai} whose parameters get fixed by the use of curvature masses of mesons after the quark one-loop vacuum correction is included in its effective potential in the $\overline{MS}$ scheme using the extended mean field approximation (e-MFA).~In the RQM model treatment of quark one-loop vacuum fluctuation,~one finds the renormalized parameters by matching the counter terms in the $\overline{MS}$ scheme with those of the on-shell scheme \cite{vkkr22, skrvkt24}.~Using the $ \mathcal{O}(\frac{1}{f^2})$  accurate results of the infrared regularized U(3) ChPT~\cite{Herpay:05, borasoyI, borasoyII, Beisert, Becher} scaling relations for the ($m_{\pi},m_{K}$) dependent expressions of the $f_{\pi},f_{K} \text{ and } M_{\eta}$,~the  parameter fixing  for the QMVT and RQM models,~have been rendered free from any ambiguity or heuristic adjustment as  the $\pi, K$ meson masses are reduced to move away from the physical point for performing  the chiral limit study.~After generating the QMVT model Columbia plots for the $m_{\sigma}=400 \text{ and } 530$ MeV,~these plots have been compared
and contrasted  with the Columbia plots generated respectively in the e-MFA:QM model FRG study  \cite{Resch} for the $m_{\sigma}=530$ MeV and the RQM model study for the $m_{\sigma}=400 \text{ and } 530$ MeV as reported very recently in the Ref.\cite{vkt25}.
 





Reducing the chiral symmetry breaking strengths $h_x$ and $h_y$ after defining the reduced pion and kaon masses $m_{\pi}^*$ and $m_{K}^*$ as $\frac{m_{\pi}^*}{m_{\pi}}= \frac{m_{K}^*}{m_{K}}=\beta$,~a direct chiral limit path is chosen such  that the $\frac{m_{\pi}^*}{m_{K}^*}$ remains equal to the ratio $\frac{m_{\pi}}{m_{K}}$ at the physical point where the ChPT ratio of the strange to light quark mass $q=\frac{2 \ m_s}{m_{ud}}$,~also gets fixed.~The temperature variations of the light and strange condensates $x \text{ and }y$,~the derivative of light condensate $\partial x/ \partial T$ and the $\sigma \text{ as well as } \pi$ meson curvature masses,~have been computed in the QMVT and RQM model for the $m_{\sigma}=530$ MeV  and their plots have been compared for different $\beta$ fractions.~The melting of the light and strange condensate is very smooth and delayed in the QMVT model as the temperature plot of the $\partial x/ \partial T$ show smaller and very smooth peak at the  pseudo-critical temperature  $T_{c}^{\chi}=162.16$ MeV when $\beta=1$ and  $T_{c}^{\chi}=158.32$ MeV when $\beta=0.5$ for the chiral crossover transition.~A relatively higher peak in the  $\partial x/ \partial T$ temperature plot of the RQM model,~gives a sharper chiral crossover transition  at about 23-24 MeV smaller transition temperature of  $T_{c}^{\chi}=139.42$ MeV when $\beta=1$ and $T_{c}^{\chi}=133.97$ MeV when $\beta=0.5$.~The peak in the RQM model $\partial x / \partial T$ temperature plot,~diverges at the $T_{c}^{\chi}=129.33$ MeV where the crossover transition ends at the second order $Z_{2}$ critical point when $\beta$ is reduced moderately from the $\beta=0.5$ to the $\beta_{cep}|_{(m_{\sigma}=530\text{ MeV}:\text{ RQM})}=0.3642 \text{ for } (m_{\pi,cep}^*,m_{K,cep}^*)=(50.26,~180.64) $.The fraction $\beta$ needs a large reduction to get the  $\beta_{cep}|_{(m_{\sigma}=530\text{ MeV}:\text{QMVT})}=0.2028 \text{ for } (m_{\pi,cep}^*,m_{K,cep}^*)=(27.99,~100.59) $ MeV in the QMVT model when the peak height of the $\partial x / \partial T$ plot diverges at larger $T_{c}^{\chi}=145.06$ MeV.~We must emphasize that the second order $Z_{2}$ critical point in the e-MFA:QM model FRG study occurs at the $\beta_{cep}|_{(m_{\sigma}=530\text{ MeV}:\text{eMFA;QM-FRG})}=\sqrt{\alpha_{c}}=0.2$ for  $(m_{\pi,cep}^*,m_{K,cep}^*)=(27.6,~99.2)$ MeV which is quite close to the QMVT model result of the present study.~Thus the quark one-loop vacuum fluctuation generates a significantly large and similar smoothing effect on the strength of chiral transition in the QMVT model as well as the e-MFA:QM model FRG study of the Ref.~\cite{Resch} and this softening effect is quite moderate in the recent RQM model study~\cite{vkt25}.~For  the smaller $\sigma$ mass $m_{\sigma}=400$ MeV,~the strength of chiral transition increases,~hence the Z(2) critical point occurs for higher fraction $\beta_{cep}|_{(m_{\sigma}=530\text{ MeV}:\text{ RQM})}=0.3795 \text{ for } (m_{\pi,cep}^*,m_{K,cep}^*)=(52.38,~188.25) $ in the RQM model and  $\beta_{cep}|_{(m_{\sigma}=530\text{ MeV}:\text{QMVT})}=0.2215 \text{ for } (m_{\pi,cep}^*,m_{K,cep}^*)=(30.57,~109.88)$ MeV in the QMVT model.

~Very close to the chiral limit when $\beta=0.00725$,~one gets huge first order gap in the QMVT and RQM model order parameters.~Note that for the smaller $\beta<\beta_{cep}$,~the  difference in the QMVT and RQM model $T_{c}^{\chi}$ decreases from about 24 MeV for the $\beta=1.0 \text{ and }0.5$ to 16.35 MeV for $\beta=0.15$ and 13.35 MeV for $\beta=0.00725$.~In the exact chiral limit $f_{\pi}=f_{K}$,~the  $m_{\eta}=0$ and the heavy $m_{\eta^{\prime}} \equiv 825$ MeV signifies the $U_A(1)$ anomaly for both the QMVT model and RQM model.~The spontaneous chiral symmetry breaking ( SCSB) for the  $m_{\sigma}=400-800$ MeV,~does not get lost in the QMVT and RQM model when  the infrared regularized U(3) ChPT scaling \cite{Herpay:05} of the  $m_{\pi},m_{K},M_{\eta}^2$ is used for the   $(m_{\pi},m_{K} \rightarrow  0)$ chiral limit study.~We point out that the $\sigma$ meson vacuum curvature mass stays the same as $m_{\sigma}=530$ MeV when the  fraction $\beta$ is reduced in the QMVT model,~whereas in the RQM model,~the $\sigma$ curvature mass $m_{\sigma,c}$ ~is lowest at the physical point and it increases successively when one moves towards the chiral limit by reducing $\beta$.~In complete contrast of the above two scenarios,~the $\sigma$ mass in the e-MFA:QM model FRG study,~registers a significant successive decrease towards the chiral limit from its physical point value of $m_{\sigma}=530$ MeV.


The second order transition turns first order at the tricritical point $m_{K}=m_{K}^{TCP}$ ($m_{s}=m_{s}^{TCP}$) on the $m_{K}(m_{s})$ axis of the Columbia plot.~The line of $Z(2)$ critical points,~intersects the  SU(3) symmetric $m_{\pi}=m_{K}$ ($m_{ud}=m_{s}$) line and terminates on the strange chiral limit line  $h_{y0}=0(m_{s}=0)$ respectively at the $m_{\pi}^{c}$ ($m_{ud}^{c}$) and  $m_{\pi}^{t}$ ($m_{ud}^t$) known as critical and terminal $\pi(\text{light quark})$ mass.~In the $m_{\pi}-m_{K}$ and $\mu-m_{K}$ plane of the QMVT model Columbia plot for the $m_{\sigma}=400$ MeV,~the computed  critical quantities in the present work  are  $( \ m_{K}^{TCP},m_{\pi}^{t},m_{\pi}^{c})=(137.2,98.15,78.45)$ MeV which give the corresponding  critical quantities in the $m_{ud}-m_{s}$ and $\mu-m_{s}$ plane as $(m_{s}^{TCP},m_{ud}^{t},m_{ud}^{c})=(8.06,2.05,1.31)$ MeV.~In the RQM model $m_{\pi}-m_{K}$ and $\mu-m_{K}$ plane of the Columbia plot when the $m_{\sigma}=400$ MeV,~one finds $( \ m_{K}^{TCP},m_{\pi}^{t},m_{\pi}^{c})=(242.6,168.39,134.16)$ MeV which give the $(m_{s}^{TCP},m_{ud}^{t},m_{ud}^{c})=(25.53,6.13,3.84)$ MeV in the  $m_{ud}-m_{s}$ and $\mu-m_{s}$ plane.~For the $m_{\sigma}=530$ MeV in the RQM model,~the critical quantities reported in the Ref.~\cite{vkt25},~are $( \ m_{K}^{TCP},m_{\pi}^{t},m_{\pi}^{c})=(235.7,160.75,128.38)$ MeV which correspond to the light-strange quark mass plane critical quantities as $(m_{s}^{TCP},m_{ud}^{t},m_{ud}^{c})=(24.08,5.58,3.51)$ MeV.~In the present QMVT model Columbia plot calculation for the $m_{\sigma}=530$ MeV,~we are finding the $( \ m_{K}^{TCP},m_{\pi}^{t},m_{\pi}^{c})=(124.7,90.03,72.12)$ MeV which give the $m_{ud}-m_{s}$ and $\mu-m_{s}$ plane critical quantities as $(m_{s}^{TCP},m_{ud}^{t},m_{ud}^{c})=(6.64,1.72,1.1)$ MeV.~Note 
that the critical light quark mass  $m_{ud}^c=1.1-1.31$ MeV is small in the QMVT model and it is $m_{ud}^c=3.51-3.84$ MeV in the RQM model whereas the recent  LQCD study in Ref.~\cite{zhang24} using  Möbius domain wall fermions  finds that $m_{ud}^c\le$4 MeV.



Comparing the QMVT and the e-MFA:QM FRG model studies,~we note that even though the fraction $\beta_{cep}$ is almost the same in both the models as $(\beta_{cep}|_{m_{\sigma}=530\text{MeV}:\text{QMVT}}=0.2028)\simeq(\beta_{cep}|_{m_{\sigma}=530\text{MeV}:\text{eMFA;QM-FRG}}\equiv0.20)$,~the extent of  first order transition regions in the QMVT model Columbia plot,~is somewhat smaller than what is found in the  e-MFA:QM model FRG Columbia plot as the critical quantities $(m_{K}^{TCP},~m_{\pi}^{t},~m_{\pi}^{c})|_{\text{QMVT}} = (124.7,~90.03,~72.12) < (m_{K}^{TCP},~m_{\pi}^{t},~m_{\pi}^{c})|_{\text{eMFA:QM-FRG}} = (169,~110,~86)$ MeV for the $m_{\sigma}=530$ MeV.~This difference is caused by different methods of approaching the chiral limit in the two studies.

In the absence of $U_A(1)$ anomaly and $m_{\sigma}=400$ MeV,~no first order region is found in the $m_{\pi}-m_{K}$ plane at $\mu=0$ in the QMVT and RQM model.~It is the effect of the quark one-loop vacuum fluctuation similar to the e-MFA:QM model FRG study~\cite{Resch}.~For the  chemical potentials $\mu<\mu_{c}$,~whatever value  the kaon mass $m_{K}$ takes in the light chiral limit $m_{\pi}=0$,~one finds a second order chiral transition everywhere in the $\mu-m_{K}$ plane for both the QMVT and RQM model.~For very high $\mu_{c}=225.4$ ($\mu_{c}=204.6$) MeV,~the first order transition appears again in the QMVT (RQM) model.~The first order region in the QMVT model above its tricritical line is smaller than the first order region that one finds above the tricritical line of the RQM model.

~The positive slope of the tricritical line in the $\mu-m_{K}$ ($\mu-m_{s}$) plane of the Columbia plot with the $U_A(1)$ anomaly,~becomes very small in the QMVT model  
for both the cases of $m_{\sigma}=400 \text{ and } 530$ MeV and becoming parallel to the $m_{K}$($m_{s}$) axis for the  $m_{K}>400$ ($m_{s}>69$) MeV,~the tricritical lines,~show good saturation trend in the QMVT model.~In contrast to the above,~the  near saturation like trend for the RQM model tricritical line present for the case of $m_{\sigma}=400$ MeV,~gets spoiled for the $m_{\sigma}=530$ MeV  tricritical line whose slope starts increasing for the $m_{K}>500$ ($m_{s}>105$) MeV.~The tricritical line for the $m_{\sigma}=530$ MeV in the QMVT model shows similar saturation trend that one finds in the tricritical line of the Fig-(3a) of the e-MFA:QM model Columbia plot in the Ref.~\cite{Resch}.~The saturation trend in the tricritical lines of the QMVT model,~sets up for about 47-52 MeV higher chemical potentials than the values of the
chemical potential around which saturation like trend appears in the tricritical lines of the RQM Model.


Our first goal of comparing the critical quantities in the chiral limit studies of the QMVT and RQM model,~has clearly established the important result that the first order regions in the $\mu-m_{K}$($\mu-m_{s}$) and $m_{\pi}-m_{K}$($m_{ud}-m_{s}$) planes of the QMVT model Columbia plot,~are  much smaller than the corresponding  RQM model first order regions.~The large first order regions in the RQM model Columbia plots are caused by the fact that the $U_{A}(1)$ anomaly strength $c$ which contains a condensate dependent part,~ gets significantly enhanced when the meson self energies due to quark loops are evaluated using the meson pole masses and parameters are fixed on-sell.~Note that the strength $c$,~being same as in the QM model,~does not change in the curvature mass based parameter fixing of the QMVT model as well as the e-MFA:QM model FRG study.The second objective of the present study led to the conclusion that  the first order regions in the  QMVT model Columbia plot are similar but moderately smaller than the extent of the first order regions reported in the $\mu-m_{K}$ and 
$m_{\pi}-m_{K}$  planes of the Columbia plot of the e-MFA:QM model FRG study for the $m_{\sigma}=530$ MeV.~The  above difference is caused by different methods of approaching the chiral limit in the two studies as the $f_{\pi},f_{K}, M_{\eta}$ and parameters,~change towards the chiral limit in the QMVT model,~whereas the $f_{\pi},~M_{\eta}$ and parameters in the e-MFA:QM-FRG study are kept fixed to their physical point value and the initial effective action is successively adjusted to the higher scales $\Lambda$ as one approaches the chiral limit. \\

\ \ \ \ \ \ \ \ \ \ \ \ \ \ {\bf {ACKNOWLEDGMENTS}} \\ \\
I would like to express my sincere gratitude towards Suraj Kumar Rai,
Akanksha Tripathi and Pooja Kumari for the careful
readings of the manuscript. I am deeply indebted and very
much tankful to Swatantra Kumar Tiwari for making the
three dimensional Columbia plots.



\bibliographystyle{apsrmp4-1}

\end{document}